\documentclass[journal]{IEEEtran}

\IEEEoverridecommandlockouts

\usepackage{cite}
\usepackage{amsmath,amssymb,amsfonts}
\usepackage{algorithmic}
\usepackage{graphicx}
\usepackage{textcomp}
\usepackage{xcolor}
\usepackage[boxruled,linesnumbered]{algorithm2e}
\usepackage{comment}
\usepackage{rotating} 
\usepackage{subfigure}
\usepackage{float}
\usepackage{amsmath}
\usepackage{svg}

\usepackage{bbding}
\usepackage{pifont}
\usepackage{wasysym}
\usepackage{amssymb}

\def\BibTeX{{\rm B\kern-.05em{\sc i\kern-.025em b}\kern-.08em
    T\kern-.1667em\lower.7ex\hbox{E}\kern-.125emX}}
\begin{document}

\title{\vspace{-30pt}On Routing Optimization in Networks with Embedded Computational Services}

\author{\IEEEauthorblockN{Lifan Mei\IEEEauthorrefmark{1},
Jinrui Gou\IEEEauthorrefmark{2}, 
Jingrui Yang\IEEEauthorrefmark{3},
Yujin Cai\IEEEauthorrefmark{4},
and
Yong Liu~\IEEEmembership{Fellow,~IEEE}\IEEEauthorrefmark{5}}\\
\IEEEauthorblockA{Department of Electrical and Computer Engineering,
New York University\\
Email: \IEEEauthorrefmark{1}lifan@nyu.edu,
\IEEEauthorrefmark{2}jr6226@nyu.edu,
\IEEEauthorrefmark{3}jy2823@nyu.edu,
\IEEEauthorrefmark{4}yc4090@nyu.edu,
\IEEEauthorrefmark{5}yongliu@nyu.edu}}

\maketitle

\begin{abstract}Modern communication networks are increasingly equipped with in-network computational capabilities and services. Routing in such networks is significantly more complicated than the traditional routing. A legitimate route for a flow not only needs to have enough communication and computation resources, but also has to conform to various application-specific routing constraints. This paper presents a comprehensive study on routing optimization problems in networks with embedded computational services. We develop a set of routing optimization models and derive low-complexity heuristic routing algorithms for diverse computation scenarios. For dynamic demands, we also develop an online routing algorithm with performance guarantees. Through evaluations over emerging applications on real topologies, we demonstrate that our models can be flexibly customized to meet the diverse routing requirements of different computation applications. Our proposed heuristic algorithms significantly outperform baseline algorithms and can achieve close-to-optimal performance in various scenarios. 
\end{abstract}


\section{Introduction} 
Modern communication networks are increasingly equipped with in-network computational capabilities and services. Software-Defined Networking (SDN)~\cite{feamster2014road}~\cite{truong2015software} technology can decouple data plane and control plane, and the routing decision can be made in a centralized fashion  rather than hop-by-hop, utilizing more information for better routing decision. Traffic flows traversing such networks are processed by different types of middle-boxes in-flight. For example, in a 5G~\cite{shafi20175g} core network, traffic from/to mobile user devices must pass through special network elements, including eNodeB, serving gateways, and packet data network gateways. To improve security and/or boost application performance, an application flow may also traverse other types of middle-boxes for application-specific processing, e.g., intrusion detection and prevention, content caching to reduce latency and network traffic, rendering of VR/AR objects to offload user devices, object detection from  video/lidar camera data acquired by autonomous vehicles, etc. In a cloud-native network, to improve performance and resilience, middle-boxes are replicated throughout the network, and can be elastically provisioned on commodity servers through Network Function Virtualization (NFV)~\cite{basta2014applying}\cite{herrera2016resource}.

Routing is a critical component of networking. The main goal of the traditional network routing is to forward user traffic to their destinations with the lowest possible delay, while maintaining the network-wide load balance and resilience. Routing in networks with embedded computational services is significantly more complicated. It has to find a path for each flow that simultaneously has sufficient bandwidth and computation resources to meet the flow's traffic and computation demands. Load balance and resilience have to be maintained on both communication links and computing nodes. To further complicate matters, application-specific computational services will introduce diverse additional routing requirements. Some application flows have to traverse multiple types of middle-boxes in certain preset orders, and the routing path may have to contain cycles. The traffic volume of a flow might increase or decrease after processing, consequently, the flow conservation law no longer holds. Some applications require the computation to be done on a single computation node, while some applications can split their computation load to multiple paths and multiple nodes to achieve the parallelization gain. How traffic and computation are split directly impacts the load balance and resilience of the whole network. The existing routing models cannot directly address these new challenges and requirements. {\it The goal of our study is to comprehensively explore the design space of routing in networks with embedded computational services. We develop a set of routing optimization models and derive low-complexity heuristic routing algorithms for diverse computation scenarios.} Towards this goal, we made the following contributions. 

\begin{enumerate}
\item For routing with non-splittable flows, we show that the problem is NP-Hard, and develop a loop-friendly mixed integer program (MIP) model to characterize the interplay between traffic routing, computation load distribution, and network delay performance. We further design a Metric-TSP type of heuristic algorithm to achieve close-to-optimal performance.

\item When flows can be arbitrarily split, we prove the equivalence between the routing with computational services problem and the regular routing problem using the {\it segment routing} idea. We develop a Linear Program (LP) routing optimization model by extending the classic Multi-Commodity-Flow (MCF) model to work with heterogeneous middle-boxes and traffic scaling resulting from the processing. The LP model can be further extended to study the joint optimization of traffic routing and computation resource provisioning. 

\item 
For come-and-go dynamic traffic demands, we convert the online routing problem into a flow packing problem, and develop a primal-dual type of online routing algorithm with performance guarantees.
 
\item We evaluate the developed models and algorithms using emerging computation applications over real network topologies. Through extensive experiments, we demonstrate that our models can be flexibly customized to meet the diverse routing 
requirements of different computation applications. Our algorithms significantly outperform baseline routing algorithms, and can achieve 
close-to-optimal performance in various scenarios.   
\end{enumerate}

\section{Related Work}
\label{sec:related}
For routing with in-network processing, there are existing studies to address various application scenarios with different assumptions. In the context of edge/fog computing, in~\cite{xu2019joint}, besides the computation allocation, they also considered mobility and privacy in the joint optimization problem. In~\cite{wang2018service}, authors focused on the service allocation problem for AR offloading.  Some papers,~\cite{rost2018charting}~\cite{rost2019virtual}~\cite{yu2018application}, focused on approximation algorithm. The authors of~\cite{rost2018charting}~\cite{rost2019virtual} used randomized rounding with linear programming relaxation as the key idea to deal with the non-splittable flows. Authors of~\cite{yu2018application} developed a fully polynomial-time approximation scheme (FPTAS) for an NP-Hard problem in IoT scenarios. In~\cite{cohen2015near}~\cite{lukovszki2018approximate}, the problem is studied without hard link capacity constraints. For the middle-box traversal order, ~\cite{valls2020online} focused on the case where the traversal order is fixed. In the~\cite{cao2017enhancing}, authors used graph layering to conform to the order of traversal. Authors of ~\cite{charikar2018multi} studied the case with arbitrary traversal order. For routing with cycles, one strategy is to calculate all paths with a certain number of cycles in advance~\cite{charikar2018multi}~\cite{chen1995restricted}~\cite{heorhiadi2016simplifying}, and then use the path-based routing formulation to find the optimal traffic routing and demand allocation. The number of candidate paths increases exponentially, and the pre-calculated paths may miss some good paths.  Among all these studies, the one that is closest to ours is~\cite{charikar2018multi}. The main assumption of their work is that flows are infinitely splittable. In real world applications, it is equally important to study non-splittable and finitely splittable flows. For infinitely splittable flows, their model assumes  universal middle-boxes while our model can handle heterogeneous  middle-boxes. Our segment routing-based formulation also makes it easy to study traffic scaling after processing and joint routing and provisioning. Table~\ref{diff} summarizes the main differences between our work and the most related studies. 

\begin{table}[!ht]\caption{Key Differences from Most Related Works}
     \label{diff}
\begin{tabular}{l|l|l|l|l|l|l|l|l|lllllllll}
\hline
{\bf Traffic Flow} &
  \multicolumn{8}{l|}{\bf Non-Splittable} &
  \multicolumn{8}{l|}{\bf Infinitely-Splittable} & 
 {\bf Scaling}  
  \\ \hline
\begin{tabular}[c]{@{}l@{}}{\bf Middle-box}\\{\bf Univ.}/{\bf Hetero.}\end{tabular} &
  \multicolumn{4}{l|}{\bf Univ.} &
  \multicolumn{4}{l|}{\bf Hetero.} &
  \multicolumn{4}{l|}{\bf Univ.} &
  \multicolumn{4}{l|}{\bf Hetero.} &
 \\ \hline
\begin{tabular}[c]{@{}l@{}}{\bf Traversal Order} \\ {\bf F}ixed or {\bf N}ot\end{tabular} &
  \multicolumn{4}{l|}{n/a} &
 % \multicolumn{2}{l|}{\bf N} &
  \multicolumn{2}{l|}{\bf F} &
  \multicolumn{2}{l|}{\bf N} &
  \multicolumn{4}{l|}{n/a} &
 % \multicolumn{2}{l|}{\bf N} &
  \multicolumn{2}{l|}{\bf F} &
  \multicolumn{2}{l|}{\bf N} &
 % \begin{tabular}[c]{@{}l@{}}Traffic \\ Scaling\end{tabular} 
  \\ \hline
\cite{charikar2018multi} &
  \multicolumn{4}{l|}{} &
 % \multicolumn{2}{l|}{} &
  \multicolumn{2}{l|}{} &
  \multicolumn{2}{l|}{} &
  \multicolumn{4}{l|}\CheckmarkBold &
 % \multicolumn{2}{l|}\CheckmarkBold &
  \multicolumn{2}{l|}{} &
  \multicolumn{2}{l|}{} &
   \\
\cite{cao2017enhancing}~\cite{valls2020online} &
  \multicolumn{4}{l|}\CheckmarkBold &
%  \multicolumn{2}{l|}\CheckmarkBold &
  \multicolumn{2}{l|}\CheckmarkBold &
  \multicolumn{2}{l|}{} &
  \multicolumn{4}{l|}{} &
%  \multicolumn{2}{l|}{} &
  \multicolumn{2}{l|}{} &
  \multicolumn{2}{l|}{} &
   \\
Our Paper &
  \multicolumn{4}{l|}\CheckmarkBold &
%  \multicolumn{2}{l|}\CheckmarkBold &
  \multicolumn{2}{l|}\CheckmarkBold &
  \multicolumn{2}{l|}\CheckmarkBold &
  \multicolumn{4}{l|}\CheckmarkBold &
%  \multicolumn{2}{l|}\CheckmarkBold &
  \multicolumn{2}{l|}\CheckmarkBold &
  \multicolumn{2}{l|}{} &
  \CheckmarkBold \\ \hline
\end{tabular}
\end{table}

\section{Routing with In-network Processing Problem}
We consider a generic network represented by a directed graph $\mathcal{G}=(\mathcal{V},\mathcal{E})$, where $\mathcal{V}$ denotes a set of nodes, and $\mathcal{E}$ denotes a set of directed links connecting the nodes. The graph $\mathcal{G}$ is assumed to be mesh-connected, having multiple paths connecting each pair of nodes. Every link $e$ is associated with bandwidth capacity of $C_e$.  The node set $\mathcal{V}$ contains standard routers and special middle-boxes attached to routers. There might be different types of middle-boxes. For a type-$r$ middle-box $z$, its processing capacity is $N_z^r$. A set of flows are to be routed and processed in the network. Each flow $d$ is characterized by its source node $s_d$, destination node $t_d$, traffic volume $h_d$, and its demand for type-$r$ processing $W_d^r$. The {\bf Routing with In-Network Processing (RINP)} problem is to find paths with sufficient bandwidth and  processing resources for all flows, subject to bandwidth/resource capacities on all links/nodes. 

There are different variations of RINP along different dimensions. 
\begin{enumerate}
\item {\it Universal vs. Heterogeneous middle-boxes:} In the  traditional networks, different types of middle-boxes are designed for specific processing tasks. The new trend is to implement middle-box functions on generic computing servers using NFV. Each computing node can be configured to process any demand. The capacity of a middle-box can be measured by its universal computation power $N^r$, and the flow processing demand can be characterized by the total computation power needed.  

\item {\it Non-splittable vs. Infinitesimal Flow/demand:} When the flow granularity is small, e.g., one flow for each user application session, splitting the flow to multiple network paths and multiple middle-boxes will be inefficient/impractical. On the other hand, in a backbone network, each flow is indeed the aggregated user traffic from city A to city B. It is therefore more flexible to split the traffic as well as the associated processing demand to multiple paths, and if a path contains multiple middle-boxes, the processing demand can be further split to utilize all available resources.  

\item {\it Constant vs. Elastic Traffic Volume:} Certain types of in-network processing will increase or decrease the traffic volume of the processed flow. For example, after an edge server rendered online game updates, the size of the rendered video stream will typically be larger than the game updates. On the other hand, after an edge server processed the Lidar data captured by an autonomous vehicle, it only needs to upload the learned representations to the cloud server, which has a volume much lower than the raw data.  

\item {\it Fixed vs. Reconfigurable Processing Capacity:} The traditional middle-boxes have fixed capacities, while the software-based virtual middle-boxes can be reconfigured on-demand to match the processing needs. The RINP problem can be more efficiently solved by jointly routing flows and provisioning middle-boxes.  
\end{enumerate}

\begin{figure}[htbp]
\centering
\subfigure[Non-splittable \label{fig:2a}] {
       \includegraphics[height=1in,width=1in]{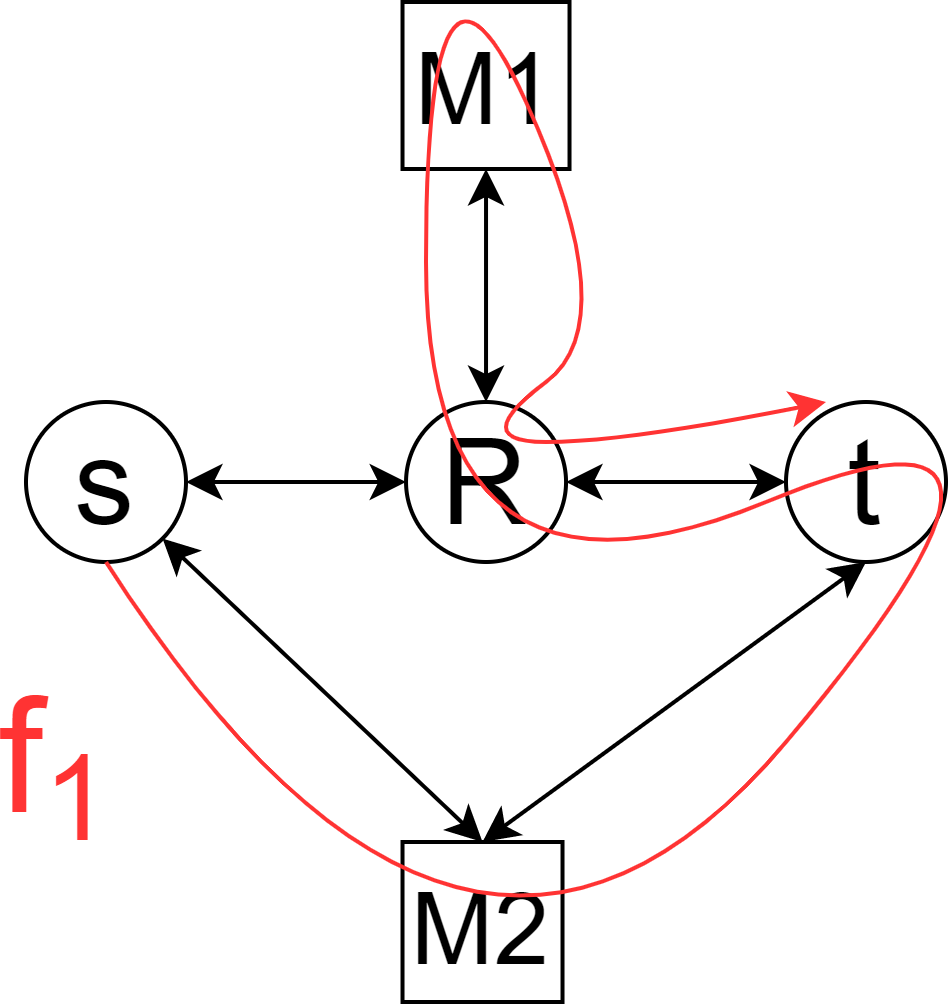}
     }
     \subfigure[Infinitely-splittable \label{fig:2b}] {
       \includegraphics[height=1in,width=1in]{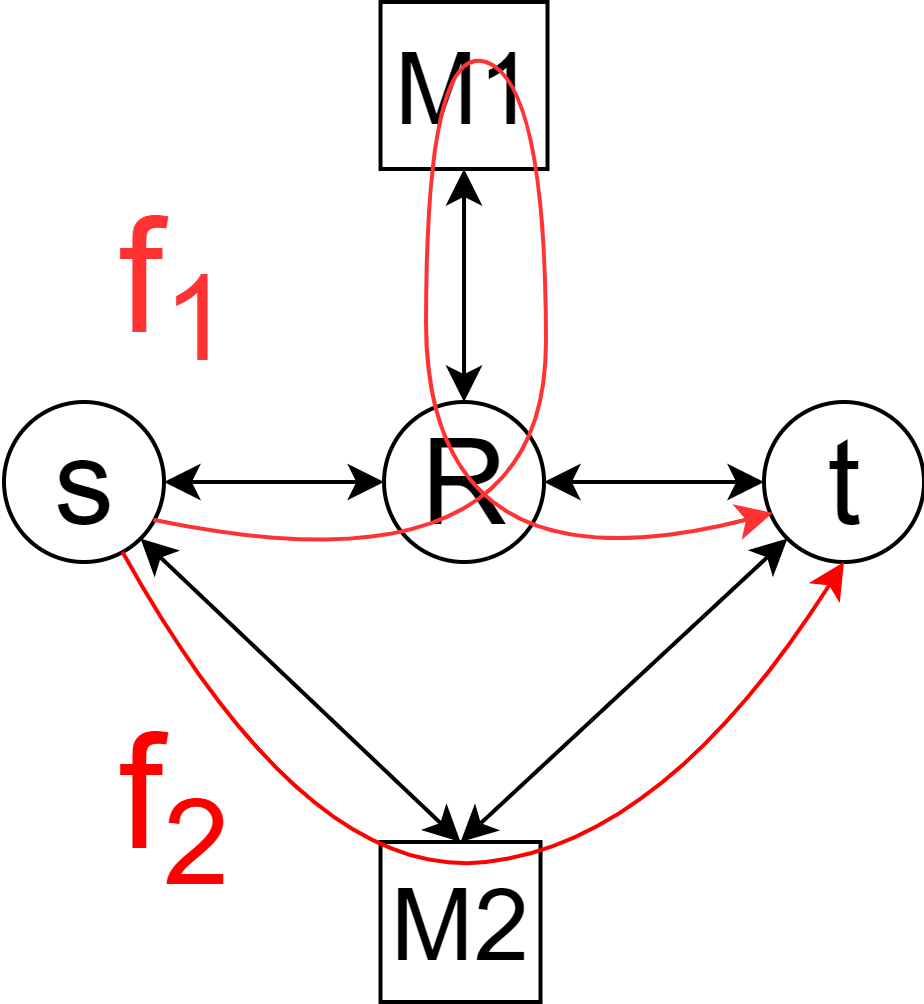}
     }
     \subfigure[Segment Routing\label{fig:2c}] {
       \includegraphics[height=1in,width=1in]{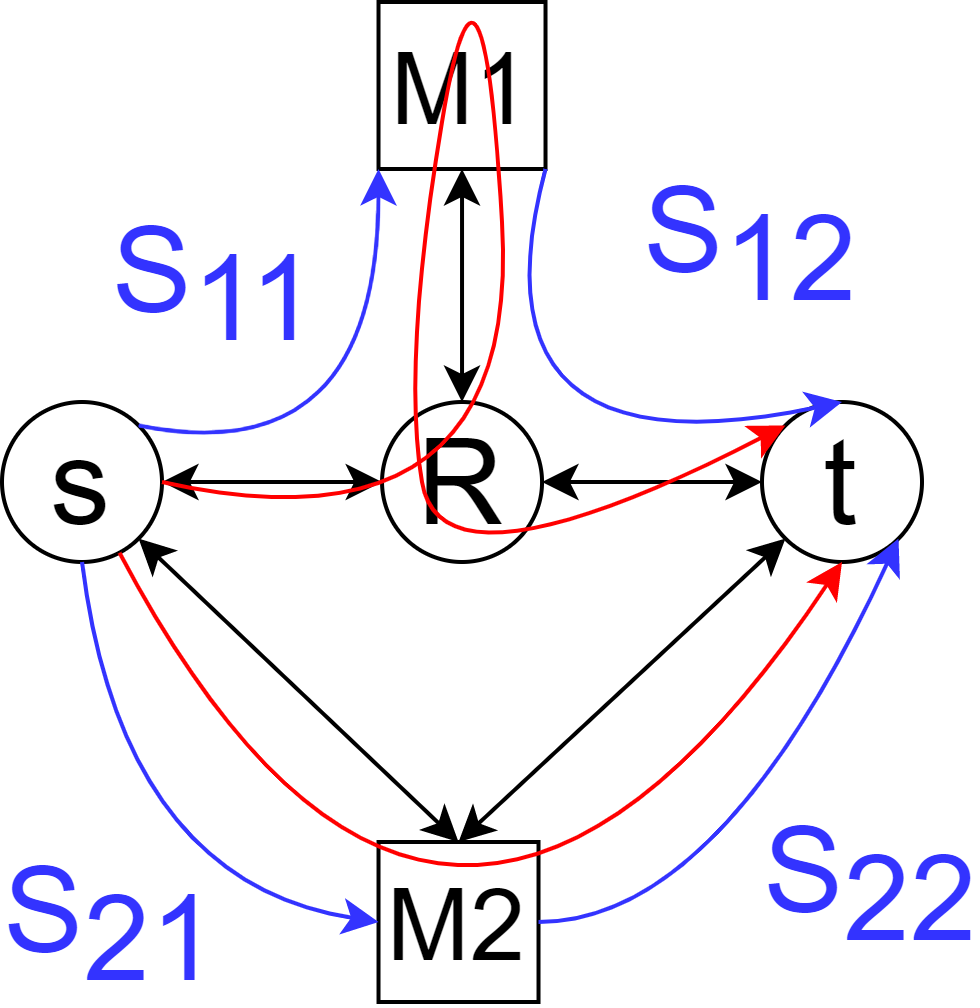}
     }

 \caption{Routing with In-Network Processing (RINP) Basic Examples \label{fig:2abc}}
\end{figure}

\begin{table}[!ht]
\centering
     \caption{Key parameters and notations}
     \label{notation}
\begin{tabular}{lll}
\hline
Symbol              & Description &  \\\hline
$V$              &  set of all nodes&  \\
$ P^r \subset V$              &  subset of computing nodes with type-$r$ resources&  \\
$E$              &  set of links in a graph&  \\
$D$              &  set of demand flows&  \\
$T$ & time horizon & \\
$a_{ev}$              &  1 if link $e$ originates from node $v$; 0 otherwise&  \\
$b_{ev}$              &  1 if link $e$ terminates at node $v$; 0 otherwise&  \\
$s_{d}$              &  source node of flow $d$&  \\
$t_{d}$              &  terminal node of flow $d$ \\
$h_{d}$         &  traffic volume of flow $d$ &  \\
$h^z_{d}$         &  traffic volume of virtual sub-flow of $d$ processed by $z$&\\ 
$W^r_{d}$     &  type-$r$ resource demand of flow $d$ &\\
% $h_d^z$        & total traffic of flow $d$ passing through compute node $z \in P$\\
$C_{e}$         &  bandwidth capacity on link $e$ &  \\
$f_{e}$              &  total traffic rate on link $e$ &  \\
$x_{ed}$              &  traffic of flow $d$ allocated on link $e$  &  \\
$x_{ed}^{zs}$       &  traffic of segment $s$ of $d$'s subflow processed by $z$ &\\ 
&allocated on link $e$&  \\ 
$u_{ed}$              &  integer, number of times flow $d$ traverses link $e$\\
$\varepsilon_{ed}$              & binary, whether or not flow $d$ traverses link $e$ \\
$N_z^{r}$         &  type-$r$ resource capacity on node $z \in P^r$ &\\
$\rho_r$            & upper bound of type-$r$ resource utilization & \\   
$N_z$              &  resource capacity on node $z$ with universal resources\\
$w_{zd}^r$              &  type-$r$ resource consumption of flow $d$ on node $z$\\
$y_{ed}^r$              & unprocessed type-$r$ demand of flow $d$ on link $e$\\

%$E$              &  set of links in a graph&  \\
%$C_{e}$         &  bandwidth capacity on link $e$ &  \\
%$D$              &  set of demands&  \\
%$h_{d}$         &  traffic volume of flow $d$ &  \\
$\tau_d, \tau_d^s, \tau_d^f$         &  duration, start time, and finish time of  $d$ &\\
$\beta_{d\tau}$ & binary, $=1$ if $\tau_d^s \le \tau \le \tau_d^f$, $=0$ otherwise &\\
$P_{d}$         &  set of candidate paths for demand $d$ &  \\
$\delta_{edp}$ & non-negative constant, the fraction of traffic of \\ & candidate path $p$ routed on link $e$\\
$y_{dp}$ & binary variable, whether demand $d$ is routed on \\ & candidate path $p \in P_d$ & \\
$D(e)$ & variable, set of demands routed on link $e$ &\\
$P(e)$ & variable, set of selected paths ($y_{dp}=1$) \\ & passing through link $e$ &\\

\hline

\end{tabular}
\end{table}

\section{RINP Optimization Models}
The optimization objective of RINP depends on the actual situation. Some popular objectives are to minimize the link/node delay, minimize the maximum link/node utilization, maximize the processed flows,  and certain combinations of them. In this paper, we use minimizing network delay as an example objective for static optimization and maximizing the processed flows for dynamic optimization. The developed formulations can be easily customized for other objectives. 

\subsection{Non-splittable Flow}
\label{sec:non-splittable}
We start with the simple case that each flow can only take a single path, i.e., non-splittable. An example is shown in Fig.~\ref{fig:2a}, there are only two middle-boxes, each with a capacity of $1$, to complete the processing demand of $2$, the flow has to take a path with cycles to pass through the two middle-boxes before reaching its destination. 
%Similarly, if the two middle-boxes are of different types, and the flow has to be processed by both types, the flow again has to take a path with cycles. In general, 
We will first show that the non-splittable  RINP problem is NP-Hard by reducing the well-known Metric Traveling Salesman Problem (TSP) to non-splittable RINP. We then develop a Mixed-Integer Program (MIP) to study the interplay between traffic routing, demand processing, and network delay optimization.

\textbf{Theorem 1}: Non-splittable RINP with constant link delays is NP-Hard. $\square$ 

{\it Proof:} 
%\begin{proof}
Given a set of nodes and the distances between them, the Traveling Salesman Problem (TSP) is to find an optimal order of traversing all the nodes with the shortest total distance. Metric-TSP (M-TSP) is a special case of TSP where the 
distances between nodes form a metric to satisfy the triangle inequality: $d(v_1,v_2) \le d(v_1,v_3) + d(v_3,v_2)$. For a given graph $G=(V,E)$ and the distance metric $d(\cdot)$, we construct a special instance of non-splittable RINP on $G$ by: 1) placing one unit of computing capacity on each node, 2) setting the propagation delay on link $(v_1, v_2)$ to $d(v_1, v_2)$, 3) creating one flow with the same source and destination node, 4) setting the flow's traffic volume to $\epsilon << C_e$ so that the congestion delay is negligible, and setting the flow's processing demand to $|V|$. Obviously, to satisfy the flow's processing demand, the flow has to visit all nodes in the graph, and to minimize the total delay RINP has to find the path with the shortest distance. The only potential discrepancy between RINP solution and the M-TSP solution is that M-TSP solution can only visit each node once while in principle RINP solution may have to visit a node multiple times, as illustrated in the example in Fig.~\ref{fig:2a}. However, due to the metric distance, we can easily show that the RINP solution for the specially constructed network can be guaranteed to be cycle-free. Suppose the RINP solution visits some nodes more than once, without loss of generality, let $k$ be the first node that is visited twice, $i$ and $j$ are the nodes visited before and after the second visit to $k$. By removing the second visit of $k$, i.e., replacing the path segment $i \rightarrow k \rightarrow j$ with  $i \rightarrow j$, the total path length can potentially be reduced due to the triangle inequality. Using this process, we can remove all the duplicate visits to get a cycle-free RINP path that has either the same length or a shorter length than the original path. This path is a cycle-free solution for M-TSP. $\blacksquare$ 
%\end{proof}

%\vspace{-7pt}
For more generic non-splittable RINP, we develop a MIP model to analytically investigate how traffic routing and demand splitting impact network-wide performance. The notations are defined in Table~\ref{notation}.  
\begin{subequations}
\begin{align}\label{MIP:1}
 & \text{\bf MIP-RINP:} \quad \underset {\{u_{ed},y^r_{ed},  w^r_{zd}\}} {\bf min} \quad 
   \begin{aligned}[t]
      &\sum_{e \in E}^{} \frac{f_{e} }{C_{e} - f_{e}}
   \end{aligned}  \\
   &\text{\it subject to} \notag \\
   &\sum_{d = 1}^{|D|} x_{ed} =   f_{e} \le C_{e},  \quad \sum_{d = 1}^{|D|} w_{zd}^r \leq \rho_r N_z^r,\label{MIP:2}\\
%   &\sum_{d = 1}^{|D|} x_{ed} \leq  C_{e}\label{MIP:3}\\
%   &\sum_{d = 1}^{|D|} y_{vd}^r \leq N_v^r\label{MIP:4}\\
   &\begin{aligned}
\sum_{e\in E} a_{ev}   u_{ed} -  \sum_{e\in E} b_{ev}   u_{ed} = \left\{\begin{matrix}
 1 & if\: \, v\: = s_d  &  \\ 
 0 & \; \; \; \; \; if\: \, v\: \neq  s_d, t_d &  \\ 
 -1 & if\: \, v\: = t_d & 
\end{matrix}\right.
\end{aligned}\label{MIP:5}\\
   & x_{ed} = u_{ed}   h_{d}\label{MIP:8}\\
   & \varepsilon_{ed} \leq  u_{ed}, \quad u_{ed} \leq  B  \varepsilon_{ed}, \quad y_{ed}^r \leq  B    \varepsilon_{ed} \label{MIP:6}\\
   &\begin{aligned}
\sum_{e\in E} a_{ev}   y^r_{ed} -  \sum_{e\in E} b_{ev}   y^r_{ed} = \left\{\begin{matrix}
 W^r_d & if\: \, v\: = s_d  &  \\
 w^r_{vd} & \; \; \; \; \; if\: \, v\: \neq  s_d, t_d &  \\ 
 0 & if\: \, v\: = t_d & 
\end{matrix}\right.
\end{aligned}\label{MIP:10}\\
& u_{ed} \ge 0 \text{ integer};  \quad \varepsilon_{ed} \text{ binary};& \label{MIP:12}\\ 
& w^r_{vd} \ge 0, w^r_{vd} =0, \forall v \notin P_r , \quad y^r_{ed} \ge 0, \label{MIP:13}&  
%
%
%
%
%   & \begin{aligned}
%\zeta_{d, r, n} + \sum_{(i, n)\in E} w_{d, r, (i, n)} 
%	\\= \sum_{(n, j)\in E}w_{d, r, (n, j)} + y_{vd}^r
%	\end{aligned}\label{MIP:10}
\end{align}
\end{subequations}
where (\ref{MIP:1}) is the total network delay modeled using the M/M/1 formula. (\ref{MIP:2}) describes the total traffic rate on a link can not exceed its bandwidth capacity, and the total type-$r$ resources consumed on a node can not be larger than its type-$r$ resource capacity discounted by the target utilization $\rho_r$. (\ref{MIP:5}) is the flow conservation for single non-splittable path. As illustrated in Fig.~\ref{fig:2a}, non-splittable flow may have to take a path with cycles. To allow a flow to traverse a link/node multiple times, routing variable $u_{ed}$ is configured to be a non-negative integer, instead of binary. On a relay node, the total number of times its outgoing links are traversed should be equal to the total number of times its incoming links are traversed. The difference between the two numbers should be $1$ for the source node, and $-1$ for the destination. (\ref{MIP:8}) calculates the total traffic carried by link $e$ for demand $d$. Since we only have a single path for $d$, whenever the path traverses link $e$, the total demand volume of $h_d$ will be added to the link. So the total traffic carried by a link is proportional to the number of times the link is traversed by the single path. Meanwhile, the resources on a node can be accessed by a flow only if the flow passes the node through links attached to it. Therefore, we want to know whether or not flow $d$ traverses link $e$, which is indicated by a binary variable 
$\varepsilon_{ed}$ in (\ref{MIP:6}).  $\varepsilon_{ed}$ is forced to take value $1$ if $u_{ed}>0$, and value $0$ if $u_{ed}=0$ (B is a large positive constant). Finally, 
(\ref{MIP:10}) is the conservation law for resources: all the resource demands exit from the source node through its outgoing links, the unsatisfied demand at the destination is zero, and on an intermediate node, the difference between the incoming unprocessed resource demand and the outgoing unprocessed resource demand is the amount of demand $w_{vd}^r$ processed on that node. Similar to $x_{ed}$ in (\ref{MIP:8}), if the path visits a link multiple times, $y_{ed}^r$ is the sum of the unprocessed demand from all the visits. (\ref{MIP:12}) are the traffic routing variables, and (\ref{MIP:13}) are the resource demand splitting variables. The two sets of variables are coupled through (\ref{MIP:6}). By replacing the convex delay function in  (\ref{MIP:1}) with a piece-wise linear function, the optimization problem becomes a mixed-integer program. 

If {\it the  processing demand is also non-splittable}, we can replace $w_{vd}^r$ by $W_d^r k_{vd}^r$, where $k_{vd}^r$ is a binary variable  indicating whether type-$r$ processing of $d$ is done on node $v$.  Then  $y^r_{ed}$ and constraint (\ref{MIP:10}) can be taken out from the formulation. The added new constraints are:  
\[\sum_v k_{vd}^r=1, \quad k_{vd}^r \le \sum_{e\in E} a_{ev} \varepsilon_{ed},\]
where the first equation requires the processing will be done on exactly one node, and the inequality dictates that a flow can only utilize processing resources on the traversed nodes. 

\subsection{Infinitely Splittable Flow}
%Based on the infinitely separable characteristics , we proposed a Linear Programming method named 1-hop method. It can be proved to be the global optimal solution.
Non-splittable flow is too limiting for large flows. The traditional Traffic Engineering (TE) of backbone networks assumes arbitrary traffic splitting and develops Multi-Commodity Flow (MCF) based convex/linear programming to optimize network design objectives. The obtained optimal routes are typically cycle-free. For RINP, if a flow can be arbitrarily split into multiple paths, the traffic and processing demand  splitting become more flexible. For the example in Fig.~\ref{fig:2b}, with traffic splitting, the flow can utilize two paths, one for traversing each middle-box. While the bottom path is cycle-free, the upper path still contains a cycle, i.e., not a simple path. Even with the infinitely splittable flow, RINP still cannot be directly solved using the MCF model. 

We address this challenge using the {\it segment routing} idea as shown in Fig.~\ref{fig:2c} . We start with universal middle-box. Flow splitting leads to both traffic and processing demand splitting among multiple paths. We assume that the processing demand allocated on a path is proportional to the traffic volume allocated on that path. If there are multiple middle-boxes on a path, the processing demand allocated on that path can be further arbitrarily split among these middle-boxes. In general, a legitimate RINP path can ``stop" at 
multiple middle-boxes to get the processing done before reaching the final destination. We define an {\it n-stop} RINP path as a path on which $n$ middle-boxes process the flow. Note, since a middle-box can simply forward traffic without processing it, an {\it n-stop} path may traverse more than $n$ middle-boxes. The single RINP path in Fig.~\ref{fig:2a} is {\it 2-stop}, while the two RINP paths in Fig.~\ref{fig:2b} are both {\it 1-stop}.

\textbf{Theorem 2}: Any {\it n-stop} RINP path can be decomposed to $n$ {\it 1-stop} RINP paths. $\square$

{\it Proof:} Let $s_d \rightsquigarrow z_1,\cdots, \rightsquigarrow z_n \rightsquigarrow,t_d$ be any {\it n-stop} RINP path. The traffic flow on this path is $f$ and the total processing demand is $w$, and the processing demand allocated to middle-box $z_i$ is $w_i$, and $\sum_{i=1}^n w_i=w$. For any $z_i$, we can generate a {\it 1-stop} RINP path that follows exactly the same route as the {\it n-stop} RINP path, but only stops at $z_i$ and gets $w_i$ amount of processing done, and all the other middle-boxes only forward the traffic. To follow the proportional demand splitting rule, the traffic allocated on this path is $fw_i/w$. It is easy to check that the $n$ {\it 1-stop} RINP paths carry the total traffic of $f$ to $t_d$, and all the processing demands of $w$ are processed. $\blacksquare$.    

\textbf{Theorem 3}: Traffic routing subproblem of RINP can be optimally solved by an equivalent Multi-Commodity-Flow routing problem for pure traffic flows. $\square$ 

{\it Proof:} We prove this by construction. 
Let $\mathcal R_d$ be a set of legitimate RINP routes for flow $d$ with total traffic volume $h_d$ and total processing demand $W_d$. And the total processing demand allocated to middle-box $z_i$ is $w_i$.  According to Theorem 2, all RINP paths in $\mathcal R_d$ can be decomposed to {\it 1-stop} RINP paths. Any {\it 1-stop} RINP path stopping at $z_i$ can be decomposed into two segments, $s_d \rightsquigarrow z_i$ and  $z_i \rightsquigarrow t_d$. The first segments of all the {\it 1-stop} RINP paths stopping at $z_i$ share the same source $s_d$ and destination $z_i$, and the total traffic flow must be $h_d  w_i/W_d$ (due to proportional demand splitting). Similarly, the second segments of all the {\it 1-stop} RINP paths stopping at $z_i$ share the same source $z_i$ and destination $t_d$, and the total traffic flow is $h_d  w_i/W_d$. In other words, the bandwidth consumed by $\mathcal R_d$ on all links can be used to carry traffic for $2|P|$ pure traffic subflows $\{s_d \rightarrow z_i, z_i \in P\}$ and $\{z_i \rightarrow t_d, z_i \in P\}$, both with traffic demand $h_d  w_i/W_d$.  
Meanwhile, it is obvious that any MCF routing solution for the $2|P|$ traffic subflows, can be used to carry traffic for $\mathcal R_d$ to enable processing $w_i$ on $z_i$. 
$\blacksquare$ 

\begin{figure}[!htb]
\centering
\includegraphics[height= 1.5in, width=1.85in]{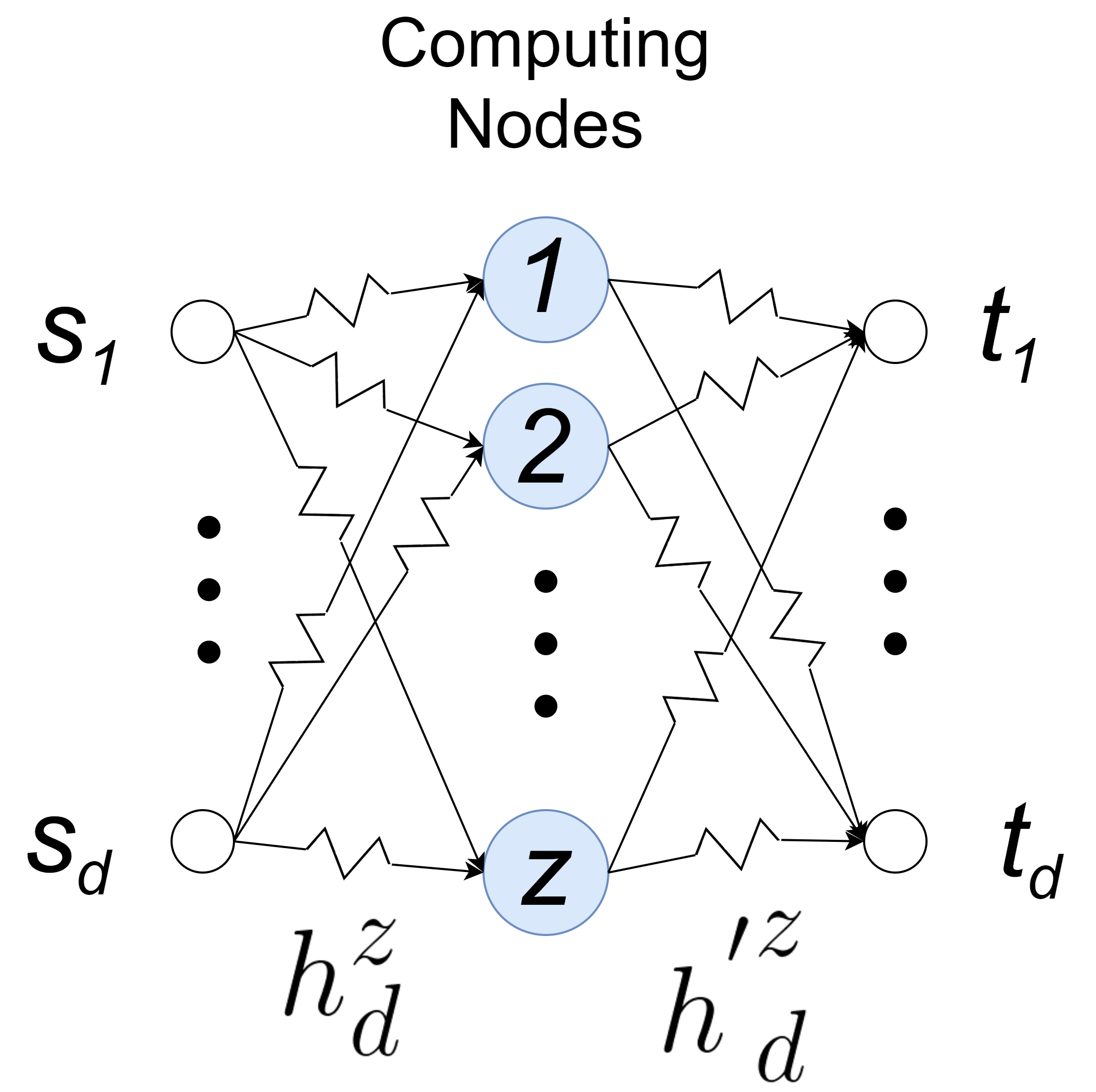}
\caption{With universal middle-boxes, any RINP flow can be implemented by two sets of regular traffic paths in the overlay graph. 
~\label{fig:1stop}}
\end{figure}

As illustrated in Fig.~\ref{fig:1stop}, based on Theorem 3, we can replace each RINP flow with two sets of traffic subflows. One set consists of subflows from the source to all middle-boxes, and the other set consists of subflows from all middle-boxes to the destination. The traffic volumes of the subflows are set to be proportional with the processing demand split among the middle-boxes. The following is an MCF-type convex/linear program for infinitely splittable RINP. 
\vspace{-0.1in}
\begin{subequations}
\begin{align}
 %  &\underset {\{u_{ed}\}} {\bf min} \quad 
  & \text{\bf SR-Infinite:} \quad \underset{\{h_d^z, x_{ed}^{z1},x_{ed}^{z2}\}}{\bf min} \quad 
   \begin{aligned}[t]
      &\sum_{e \in E}^{} \frac{f_{e} }{C_{e} - f_{e}}
   \end{aligned}  \label{HOP:1}\\
   &\text{\it subject to} \notag \\
   &\sum_{z \in P} h_d^z = h_{d}, \quad \sum_{d \in D} \frac {h_d^z}{h_d} W_d \leq  \rho N_{z} \label{HOP:3}\\
%   &\sum_{z \in P} \omega_{d,z} = h_{d}\label{HOP:4}\\
%&\upsilon_{d,z} = \omega_{d, z}\label{HOP:5}\\
%&\sum_{d}^{} \upsilon_{d,z} \leq  N_{z}\label{HOP:6}\\
&\sum_{d \in D} \sum_{z \in P} (x_{ed}^{z1}+x_{ed}^{z2} ) \leq  C_{e}\label{HOP:7}\\
%&\begin{aligned}
%\sum_{e}^{}  a_{ev}   x_{ed}^{zs}  - \sum_{e}^{}  b_{ev}   x_{ed}^{zs} =  0 \label{HOP:8}\\ \; where\; n \neq s_{d}, t_{d}, z 
%\end{aligned}\\
&\begin{aligned}
\sum_{e\in E}^{}  a_{ev}   x_{ed}^{z1}  - \sum_{e\in E}^{}  b_{ev}   x_{ed}^{z1} =   \left\{\begin{matrix}
h_d^z  &     if \; v =s_d \\ 
 -h_d^z &    if \; v =z\\
 0 & otherwise  
\end{matrix}\right.
\end{aligned}\label{HOP:11}\\
&\begin{aligned}
\sum_{e\in E}^{}  a_{ev}   x_{ed}^{z2}  - \sum_{e\in E}^{}  b_{ev}   x_{ed}^{z2} =   \left\{\begin{matrix}
h_d^z   &     if \; v = z\\ 
-h_d^z &    if \; v = t_{d}\\
 0 & otherwise    
\end{matrix}\right.
\end{aligned}\label{HOP:12}\\
& x_{ed}^{z1}\ge 0 ,x_{ed}^{z2}\ge 0,  h_d^z \ge 0,
\label{HOP:14}
\end{align}
\end{subequations}
where (\ref{HOP:3}) is the allocation of processing demand/traffic among all computing nodes. Here we assume the resource demand splitting is proportional to the traffic splitting. (\ref{HOP:7}) dictates that the traffic of the first and second segments share link capacity, 
(\ref{HOP:11}) is the flow conservation for the first segment demand from $s_d$ to $z$, while (\ref{HOP:12}) is the flow conservation 
for the second segment demand from $z$ to $t_d$. Both segments have an identical volume of $h_d^z$. If the traffic volume increases/decreases after being processed, we only need to change ${h_{}^{'}}_{d}^{z}$ to $\phi_d h_d^z$ in (\ref{HOP:12}), where  $\phi_d$ is the ratio of the traffic volume after processing over the original volume.  After being processed by computing node $z$, all traffic going to the same destination can be aggregated and routed together. The total volume of the aggregated demand from $z$ to $v$ is $\sum_{d:t_d=v} h_d^z$. To further reduce the number of routing variables and routing constraints in the formulation, we can replace flow-based routing variables with {\it destination-based routing variables}. Let $\eta_{ev}$ be the amount of post-processing traffic (regardless of the processing node) destined to node $v$ (regardless of the source), the aggregated second-segment routing constraints on any node $v' \in V$ can be rewritten as: 
\begin{equation}
\label{eq:agg_2nd}
\begin{aligned}
\sum_{e\in E}  a_{ev'}   \eta_{ev}  - \sum_{e\in E}  b_{ev'} \eta_{ev}  =   \left\{\begin{matrix}
 \sum_{d:t_d=v} h_d^{v'}   &     if \; v' \in P\\ 
\sum_{d:t_d=v} h_d &    if \; v'= v\\
 0 & otherwise,    
\end{matrix}\right.
\end{aligned}\\
\end{equation}
where the left-hand side is the difference between the outgoing traffic destined to $v$ and incoming traffic destined to $v$, the right-hand side means if $v'$ is one of the computing node, the difference is simply the total post-processing traffic from $v'$ to $v$, if $v'$ is the destination itself, then the difference is all post-processing traffic to $v$, for other relay nodes, the difference should be zero. The number of routing variables is reduced from $|V|^2|P||E|$ in (\ref{HOP:12}) to $|V||E|$ here. The number of routing constraints is also reduced from $|V|^3|P|$ to $|V|^2$. Both are reduced by a factor of $|V||P|$. By replacing (\ref{HOP:12}) with (\ref{eq:agg_2nd}), and replacing  (\ref{HOP:7}) with 
\[\sum_{d \in D} \sum_{z \in P} x_{ed}^{z1}+\sum_{v \in V} \eta_{ev} \leq  C_{e},\]
we have a more compact optimization problem.

\vspace{-10pt}\subsection{Joint Routing and Resource Provisioning Problem}
The routing optimization models can be easily extended to study the joint optimization of traffic routing and middle-box provisioning by  making the middle-box capacity $N_z^r$ in (\ref{MIP:2}) and  $N_z$ in (\ref{HOP:3}) variables under some total resource budget constraint. 
%Other than limiting the resource utilization, one can also put the load-sensitive processing delay into the optimization objective. 
% \[\]
Flexible resource provisioning can make the routing job easier, and can play an important role in network failure recovery. We will demonstrate this through case studies in Section~\ref{Placement}.

\vspace{-8pt}\subsection{Heterogeneous Middle-boxes} 
When middle-boxes are heterogeneous, each flow can have multiple types of processing demands.  

\subsubsection{Non-splittable Flow}
The MIP optimization model in Sec.~\ref{sec:non-splittable} has already considered different types of demands. In the optimal solution, 
the traversal order of different types of middle-boxes can be arbitrary. This may not be acceptable for certain application scenarios. For example, in cellular core networks, there are predefined orders for middle-box traversal, e.g., mobile traffic has to first go through a firewall before being routed to a load balancer. RINP with predefined middle-box  traversal order was studied in~\cite{cao2017enhancing} using graph layering.

\subsubsection{Infinitely Splittable Flow}
\begin{figure}
\centering
\includegraphics[height= 1.5in, width=2.3in]{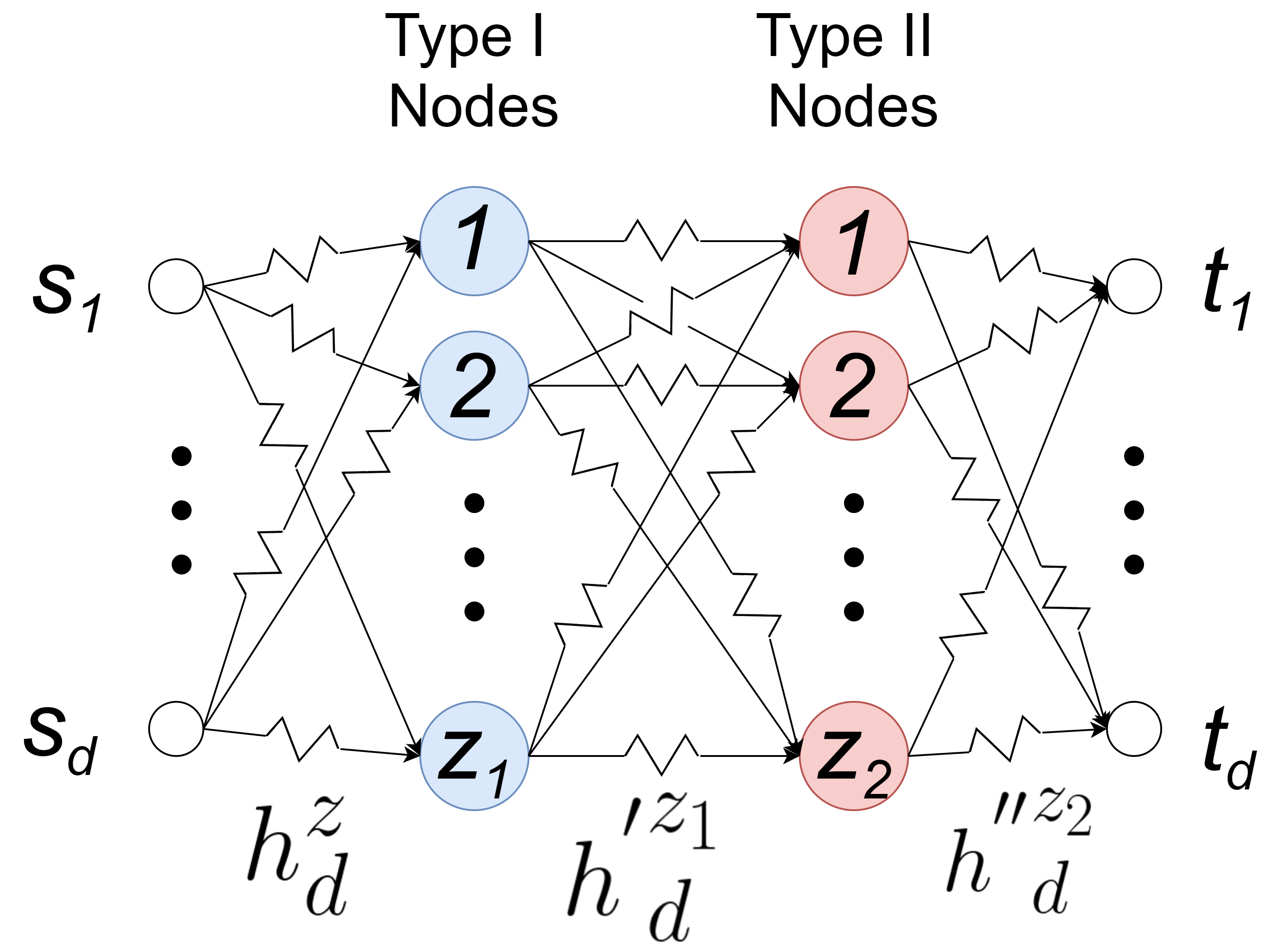}
\caption{With two types of middle-boxes, any RINP flow can be implemented by three sets of regular traffic paths in the overlay graph.~\label{fig:2stop}}
\end{figure}

When the traversal order of heterogeneous middle-boxes is predefined, the LP formulation for homogeneous middle-box can be extended to study heterogeneous middle-boxes. If there are $k$ types of middle-boxes, let the middle-box index $i$ represent its order of traversal. Extending the 2-segment routing idea, each RINP path now consists of $k+1$ segments, $s_d \rightsquigarrow z^{(1)} \cdots \rightsquigarrow z^{(k)} \rightsquigarrow t_d$, where  $z^{(i)}$ is some type-$i$ middle-box. Similar to Theorem 3, each RINP demand can be replaced by $k+1$ sets of pure traffic subflows, one set for each segment.   The number of subflows at segment $k$ is $|P_{k-1}||P_{k}|$.  An example of two-types of middle-boxes is illustrated in Fig.~\ref{fig:2stop}. MCF linear/convex program model can be established in a similar fashion to  (\ref{HOP:1}). \vspace{-3pt}Due to the space limit, we skip the detailed model here and refer interested users to our technical report~\cite{Lifan_Report22}. 

\vspace{-5pt}
\section{Fast Heuristic Routing Algorithms}

\subsection{Heuristic for Non-splittable Flow}
Even though the non-splittable version of RINP is NP-Hard, we can still leverage the solution of the infinitely splittable RINP to develop a close-to-optimal single path solution. More specifically, we relax the non-splittable requirement and get the optimal infinitely splittable solution by solving the LP defined in~(\ref{HOP:1}). While the paths cannot be used, the processing demand allocation, namely $\{\frac {h_d^z} {h_d} W_d\}$ can be used for the single path solution. For all computing nodes with non-zero processing demand allocation, we then find with a {\it close-to-shortest} path to traverse them using an approximation algorithm solving the Metric-TSP problem, as illustrated  in Algorithm~\ref{H:1}: 

1. We sort all flows by their demand volumes;

2. We solve the LP of segment routing with the infinitely splittable flow, denoted by $SR$ in line 3, to get a set of computing nodes $\mathcal Z_d$ to be used by each flow $d$, and the demand allocation on each computing node $\{\frac {h_d^z} {h_d} W_d, z \in \mathcal Z_d\}$;

3. Process all flows in the decreasing order of volume. For each flow $d$, find the path with enough capacity by removing links with a capacity less than its volume $h_d$ (line 6). Generate an overlay graph $G^z$ consisting of the source, destination, and all the used computing nodes for demand $d$ (line 7). The distance between two overlay nodes is the current shortest path in the underlay network (considering the congestion delay) (line 8). 

4. Find the shortest path in the overlay graph $G^z$ to traverse all computing nodes $\mathcal Z_d$ using M-TSP approximation algorithm. (line 9). We use Christofides algorithm~\cite{christofides1976worst} here. 
 
5. Map the overlay path back to the underlay network (line 10), take the bandwidth consumed by flow $d$ out of the underlay network (line 11), and process the next flow.

6. Evaluate the network delay and return.

\begin{algorithm}[!] 
\SetAlgoLined
\label{H:1}
\begin{algorithmic}[1]
\STATE \textbf{Input} : Underlying Network $G= (V, E)$, Set of Computation Flows $D$
\STATE $D = Sort(D, h_d)$
\STATE $\{\mathcal Z_d, \forall d \in D\}  \Leftarrow SR(G, D)$
\FOR{$d$ in $D$}                    
\STATE $G' = G$
\STATE $G'.remove(\left \{e| e.remain\_capacity \leq h_d \right \})$
\STATE $G^{z}.V = \mathcal Z_d \bigcup \left \{ s_d, t_d \right\}  $
\STATE $d^z (v,v')$ $=$ $Shortest\_Path(G', v,v')$, \\ $\forall v, v' \in G^{z}.V$
\STATE $\mathcal P^{z}(s_d,t_d)= \text{M-TSP}(G^{z}, s_d, t_d)$
\STATE $ \mathcal P(s_d,t_d)= Recon(G, \mathcal P^{z}(s_d,t_d))$
\STATE $G.Update(\mathcal P(s_d,t_d), h_d, \mathcal Z_d)$
\ENDFOR
\RETURN $\sum_{e \in E}^{} e.delay$
\end{algorithmic}
\caption{ \textsc{Heuristic: SR-TSP} }
\end{algorithm}

\vspace{-10pt}\subsection{Finitely Splittable Flow}
Real situations may lie between the infinitely splittable flow and the non-splittable flow. While non-splittable is too rigid, splitting flow to too many paths will introduce much overhead for implementation, such as too many flow entries in the routing table. Finitely splittable means that the maximum number $k$ of paths each flow can be split to is controlled. One way to get a {\it k-split} solution is to evenly split traffic and computation demands of a flow equally and get $k$ subflows, each of which can be treated as an independent flow and obtain the RINP solution using the MIP formulation in (\ref{MIP:1}). We call this scheme {\bf MIP-K}. However, such an approach has to work with a set of $k|D|$ subflows. It is impractical to solve MIP for any reasonably large network. 

\subsubsection{Heuristic}
To address this scalability issue, we came up with a fast heuristic algorithm. Motivated by that Segment Routing can lead to the optimal solution for the infinitely splittable flow case, our heuristic algorithm also follows the segment routing framework. We first equally divide each flow into $k$ subflows, then iteratively find the shortest two-segment path for each subflow, and update the computation and bandwidth resource capacities. If two subflows of the same original flow share the same path, they will be merged them back into a larger subflow.  The pseudo-code is in Algorithm~\ref{H:2}. For each demand list, we sort the subflows in descending order of their volumes. In each  iteration, find all the available computing nodes, $\mathcal Z_d$, whose available computation resource is more than the current computation demand. For every 2-segment path through $z \in \mathcal Z_d$, find the shortest (smallest delay) path from $s_d$ to some computing node $z$, and then to $t_d$. In the pseudo-code, $BP$ and $BPC$ denote the best path, and the best path cost. $SP$ and $SPC$ are the functions for computing the shortest path and shortest path cost in the current graph (considering the link congestion delay). %After the path for a subflow is found, the graph will be updated to take out the bandwidth and computation resources it consumed. 

\begin{algorithm}[!] 
\SetAlgoLined
\label{H:2}
\begin{algorithmic}[1]
\STATE \textbf{Input} : $G= (V, E), D$
\STATE Equally split each flow in $D$ to k subflows, get a new demand set $D^k$
\STATE $D^k = Sort(D^k, h_d)$
\FOR{$d$ in $D$}  
\STATE $\mathcal Z_d =  \left \{ v\in G.V | v.remain\_comp > W_d \right \}$
\STATE $BPC = +\infty $
    \FOR{$z$ in $\mathcal Z_d:$}
        \IF{$BPC > SPC(s_d, z) + SPC(z, t_d)$}
            \STATE $BPC = SPC(s_d, z) + SPC(z, t_d)$
            \STATE $BP = SP(s_d, z) \bigcup SP(z, t_d)$
        \ENDIF
    \ENDFOR
    \STATE $G.update(Path, h_d, W_d, BP)$
    \ENDFOR
\STATE Merge subflows of the same flow $d \in D$
\STATE \Return $\sum_{e \in E}^{} e.delay$

\end{algorithmic}
\caption{ \textsc{SR-Iteration for k-split} }
\end{algorithm}

%\begin{algorithm} \label{H:2}
%\SetAlgoLined
%\textbf{Input} : $G= (V, E), D$\;
%Equally split each flow in $D$ to k subflows, get a new demand set %$D^k$\\ 
%$D^k = Sort(D^k, h_d)$\\
%$for \; d \; in \; D^k:$\\
%\hskip 2em $\mathcal Z_d =  \left \{ v\in G.V | v.remain\_comp > %W_d \right \}$  \\
%\hskip 2em $BPC = +\infty $\\
%\hskip 2em $for \; z \; in \; \mathcal Z_d:$\\
%\hskip 4em $if \; BPC > SPC(s_d, z) + SPC(z, t_d):$\\
%\hskip 6em          $BPC = SPC(s_d, z) + SPC(z, t_d)$ \\
%\hskip 6em          $BP = SP(s_d, z) \bigcup SP(z, t_d)$ \\
%\hskip 2em $G.update(Path, h_d, W_d, BP)$\\

%Merge subflows of the same flow $d \in D$;\\
%$Return \; \sum_{e \in E}^{} e.delay$\\
% \caption{ \textsc{SR-Iteration for k-split} }
%\end{algorithm}

\vspace{-10pt}\section{Online Routing Algorithm with Performance Guarantee}
 The previous formulations assume the application demands are known, and can be used for routing optimization in long-term traffic engineering. In practice, application flows come and go with a finite duration. Whenever a new application flow joins the network, the online routing algorithm has to find a feasible routing path for it without knowing the future application flows.

\vspace{-10pt}\subsection{Online Routing Optimization Models}
Similar to other online routing studies, e.g.~\cite{buchbinder2009design,cao2017enhancing}, we want to accept as many flows as possible over time. The objective of online routing optimization is to maximize the total accepted flows, while complying with the capacity constraints on computation nodes and communication links.

\begin{figure}
\centering
\includegraphics[height= 1.2in, width=3.5in]{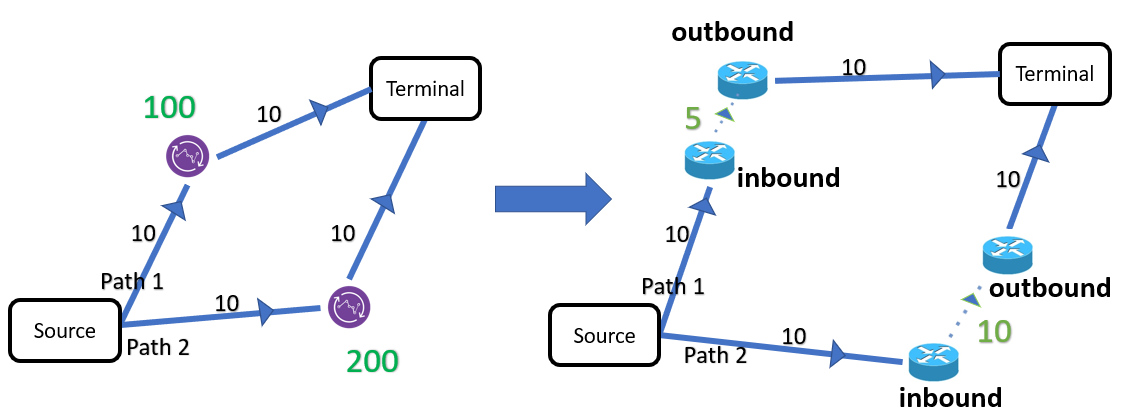}
\caption{Model Computing Node Capacity using Virtual Link~\label{fig:conver}}
\end{figure}

\textbf{Node Capacity Constraint Conversion}:
To simplify the problem, we convert computation node capacity constraints into virtual link capacity constraints. More specifically, each computing node is split into two virtual routing nodes: the inbound node connects to the computing node's  inbound links, and the outbound node connects to the computing node's outbound links. A virtual link is introduced from the inbound node to the outbound node. The bandwidth capacity of 
the virtual link is set as the computing node capacity divided by the resource demand of per unit traffic. Consequently, if the  traffic flow traversing the virtual link is bounded by the link capacity, the resource demand of the flow is bounded by the computing node capacity. In Fig.~\ref{fig:conver}, the purple nodes are routers with computing resources capacity of 100 and 200, respectively. Each computing node is split into two copies in the extended virtual graph. Solid lines and dotted lines mark physical and virtual links, respectively. The demand is from $Source$ to $Terminal$, and its traffic volume is 8 units, and needs 160 units of computing resources. The resource demand of each traffic unit is $160/8 = 20$. The capacity of virtual links introduced for the two computing nodes is $100/20 = 5$ and $200/20 = 10$, respectively. After such conversion, the bottleneck of the upper path is $5$, and only the lower path is feasible for the demand.

\begin{figure}
\centering
\includegraphics[height= 1.2in, width=3.5in]{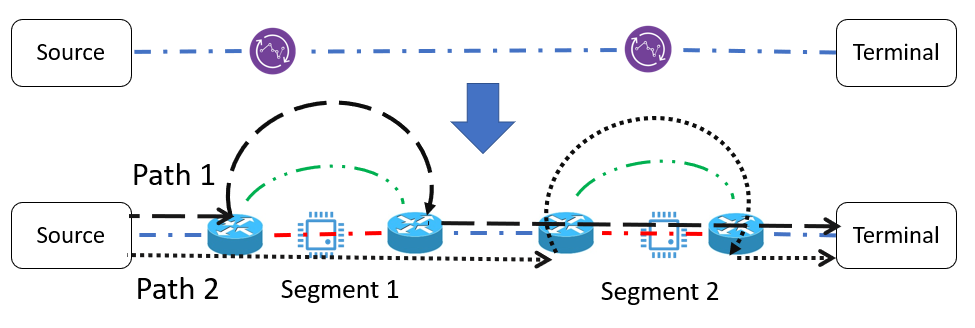}
\caption{Generating Candidate Path with Computing Resources~\label{fig:Path_Gen}}
\end{figure}

\textbf{Candidate Path with Computation Resources}:
After the conversion at the previous step, not all the paths complying with the link capacity constraints are feasible. For example, if a flow does not traverse any virtual link, its computing resource demand will not be satisfied. To address this issue, we introduce two virtual links from the inbound copy to the outbound copy of each computing node: if a flow traverses the red link, it consumes the computing resources on the node; if a flow traverses the green link, it is only forwarded by the node without consuming any computing resources. The red virtual link capacity is set according to the computing node capacity, and the green link capacity can be set to a large value so that it is never a bottleneck. By requiring each candidate path to contain exactly one red link, we can make sure that a flow will be processed correctly when routed on a candidate path. Fig.~\ref{fig:Path_Gen} shows an example of generating candidate paths in a network with two computing nodes. After removing the paths with either no red link, or two red links, there are two valid paths:  $Path\:1$ uses the green link for Segment 1 and the red link for Segment 2, and $Path\:2$ uses the red link for Segment 1 and the green link for Segment 2.

After the previous conversion and candidate path generation, we now present the routing model to maximize the total accepted demands:
\begin{subequations}
\begin{align}\label{Routing}
   &\underset {y_{dp}} {\bf max} \sum_{d} \tau_d h_d \sum_{p \in P_d} y_{dp}\\ 
   &\text{\it subject to} \notag \\
   & \sum_{p \in P_d} y_{dp} \le 1, \quad \forall d \in D,\\
   & \sum_{d} h_d \beta_{dt} \sum_{p \in P_d} \delta_{edp} y_{dp} \le C_e, \quad \forall e \in E, \forall t \in T,  
\end{align}   
\end{subequations}
where $P_d$ is the set of candidate paths for flow $d$, and $y_{dp}$ is a binary variable indicating whether $d$ is routed on path $p$. Other notations are defined in Table~\ref{notation}.

By introducing Lagrange multipliers $(z_d, d \in D, x_{et}, e \in E, t \in T)$ for the constraints, the Lagrangian function $L(\{z_d\}, \{x_{et}\})$ is: 
\begin{gather}
\nonumber \sum_{d} \tau_d h_d \sum_{p \in P_d} y_{dp}-\sum_d z_d\left (\sum_{p \in P_d} y_{dp} -1\right) \\ -\nonumber \sum_{e,t} x_{et} \left(\sum_{d} h_d \beta_{dt} \sum_{p \in P_d} \delta_{edp} y_{dp} - C_e \right),
\end{gather}

The corresponding dual optimization problem becomes:
\begin{subequations}
\begin{align}\label{Covering}
   &\underset {z_d,x_{et}} {\bf min} \sum_{d} z_d + \sum_{e,t} C_e x_{et}\\ 
   &\text{\it subject to} \notag \\
   & z_d+\sum_{e,t} x_{et}h_d \delta_{edp} \beta_{dt} \ge \tau_d h_d, \quad \forall d \in D, \forall p \in P_d
\end{align}   
\end{subequations}

Due to the strong duality theorem, any feasible solution of the dual problem is the upper bound of the optimal solution of the routing problem. 

\vspace{-12pt}\subsection{Online Primal-Dual Algorithm}
While the optimization problem can be solved offline, we need an online routing algorithm to route each flow as it arrives without knowing the future flows. It is expected that the total accepted flow volume by the online routing algorithm is lower than the offline optimal solution. Following the general framework of the online approximation algorithm~\cite{buchbinder2009design}, we develop the following online routing algorithm with a guaranteed ratio of accepted traffic over the offline optimum.
\begin{algorithm}[!] 
\SetAlgoLined
\label{Algo3}
\begin{algorithmic}[1]
\STATE \textbf{Input} : $G= (V, E)$
\STATE \textbf{Initialize:} $x_{et}=0, z_d=0$, $y_{dp}=0$
\STATE  Whenever a new demand $d$ is introduced, 
    \IF{there is a path $p \in P_d$, so that\\
        \STATE \label{eq:condtion1} \  \ \ $\sum_{e,t} x_{et} \delta_{edp} \beta_{dt} < \tau_d $}
    \STATE Set $y_{dp}=1$, $z_d=\tau_d h_d$
    \FOR{each link $e$ on path $p$} 
    \STATE $ D(e)=D(e)+\{d\} $
    \STATE \label{eq:update1} $ P(e)=P(e)+\{p\} $
    \STATE \label{eq:updateline10}       $x_{et} = x_{et} \left( 1+ \frac {h_d \delta_{edp}} {C_e}\right) + \frac {h_d \delta_{edp}} {C_e \Delta_{dp}} $,
    \STATE $where$ $\Delta_{dp} \triangleq \sum_{e \in E} {\delta_{edp}}$
    \ENDFOR
    \ELSE  \STATE Reject the demand $d$
    \ENDIF 
\end{algorithmic}
\caption{ \textsc{Online Routing Algorithm} }
\end{algorithm}

In Algorithm~\ref{Algo3}, when a new demand $d$ is introduced, we first check whether there exists a path that satisfies the dual constraint. If so, that path is picked, and update the dual variable $x_{et}$ and decision variables, $D(e)$ and $P(e)$, accordingly. If not, reject the demand $d$. Due to Line~\ref{eq:condtion1} in Algorithm~\ref{Algo3}, at any step of the update, before the update, $x_{et}(start) \le  \gamma_e \triangleq \underset{p \in P(e)} {\max} \frac {\tau_d} {\delta_{edp}}$. After the update, assuming $h_d \delta_{edp} < C_e$, we have 
\begin{equation}
\label{eq:bound1}
x_{et} \le 2 \gamma_e+1.
\end{equation}

%\end{itemize}

\vspace{-10pt}\subsection{Performance Guarantee}

 \textbf{Theorem 4}: The accepted demands by the online routing algorithm is $ \ge \frac 1 3$ of the offline optimal solution. $\square$ \\
\\
{\it Proof:}
Obviously, the $(z_d, x_{et})$ generated by the algorithm is a feasible solution for the dual problem ($z_d=\tau_d h_d$). 
After each demand is introduced, the increase in the routing objective is simply $\tau_d h_d$, resulting from $y_{dp}$ increases from 0 to 1;
The increase in the dual objective is:  
%\begin{equation}
% \tau_d h_d + \sum_{e,t} C_e \Delta x_{e,t} \delta_{edp} \beta_{dt}=  2 \tau_d h_d + \sum_{e,t} x_{et} h_d \delta_{edp} \beta_{dt} \le 3 \tau_d h_d,
%\end{equation}   

\begin{equation}
 \tau_d h_d + \sum_{e,t} C_e \Delta x_{et} =  2 \tau_d h_d + \sum_{e,t} x_{et} h_d \delta_{edp} \beta_{dt} \le 3 \tau_d h_d,
\end{equation}

where the last inequality is due to Line~\ref{eq:condtion1} in Algorithm~\ref{Algo3}. After all iterations, the objective value of the dual solution is less than three times the objective of the routing solution. Since dual feasible solution is the upper bound of the accepted demands of the offline optimal routing solution, so the accepted demands of the online routing solution are no less than $\frac 1 3$ of the offline optimal routing solution. $\blacksquare$ 
\\

\textbf{Theorem 5}: Upper bound for the link capacity violation of online routing algorithm is $W_e \log (\Delta_e (2 \gamma_e+1)+1)$  $\square$ \\
\\
{\it Proof:}
It is clear that for any $h_d\ge 1$ (Taylor Expansion), 
\[1+ \frac {h_d \delta_{edp}} {C_e} \le \left ( 1 + \frac 1 {C_e}\right )^{h_d \delta_{edp}}.\]
We can choose $W_e$ so that 
\begin{equation}
\label{eq:condition2}
1+ \frac {h_d \delta_{edp}} {C_e} \ge  \left ( 1 + \frac 1 { W_e C_e}\right )^{h_d \delta_{edp}}, \forall p \in P(e).
\end{equation}
In other words,
 \begin{align}
 W_e &\triangleq \underset{p \in P(e)} {\max} \frac 1 {C_e\left ( \left(1+\frac {h_d \delta_{edp}} {C_e}\right)^{\frac 1 {h_d \delta_{edp}}} -1 \right)}\\
 &= \frac 1 {C_e\left ( \left(1+\frac {\alpha_e} {C_e}\right)^{\frac 1 {\alpha_e}} -1 \right)}, 
\end{align}
where $\alpha_e \triangleq \underset{p \in P(e)} {\max} h_d \delta_{edp}$, and the last equality holds due to function $(1+\frac x c)^{1/x}$ is a decreasing function of $x$. $\square$

When $h_d<1$, due to Bernoulli's Inequality, (\ref{eq:condition2}) holds naturally by setting $W_e = 1$. Now define another sequence $x'_{e,t}$ and update it synchronously with $x_{e,t}$ as:
\begin{align}\label{eq:update_eq111}
\nonumber x'_{et} = x'_{et} \left( 1+ \frac 1 {W_e C_e}\right)^{h_d \delta_{edp}} 
\\ 
+ \frac 1 {\Delta_e} \left( \left( 1+ \frac 1 {W_e C_e}\right)^{h_d \delta_{edp}} -1 \right ), 
\end{align}
where $\Delta_e \triangleq \underset{p \in P(e)} {\max} \Delta_{dp}$.
Comparing Line~\ref{eq:updateline10} in Algorithm~\ref{Algo3} with (\ref{eq:update_eq111}), due to (\ref{eq:condition2}), both the multiplicative increase factor (the first term) and the  additive increase (the second term) of $x_{et}$ are larger than $x'_{et}$. So we can conclude $x_{et} > x'_{et}$.

\noindent Meanwhile, (\ref{eq:update_eq111}) can be transformed into: 
 \[x'_{et}(end) +  \frac 1 {\Delta_e} =  (x'_{et}(start)  + \frac 1 {\Delta_e}) \left( 1+ \frac 1 {W_e C_e}\right)^{h_d \delta_{edp}},\]
 together with $x'_{et}(0)=0$, at the end of the online algorithm, 
 \[x'_{et}= \frac 1 {\Delta_e} \left(\left( 1+ \frac 1 {W_e C_e}\right)^{\sum_{p \in P(e)} h_d \delta_{edp}}-1 \right)\]
 Then we have for $C_e >>1$: 
 \[\sum_{p \in P(e)} h_d \delta_{edp} = \frac {\log (\Delta_e x'_{et}+1)} {\log (1+\frac 1 {C_e W_e})} \approx C_e W_e \log (\Delta_e x'_{et}+1).\] 
 
Finally, due to $x_{et} > x'_{et}$ and (\ref{eq:bound1}), we have 
\begin{equation}
\label{eq:bound2}
\sum_{p \in P(e)} h_d \delta_{edp} \le C_e W_e \log (\Delta_e (2 \gamma_e+1)+1).
\end{equation}
In other words, the capacity violation factor on any link is bounded by $W_e \log (\Delta_e (2 \gamma_e+1)+1)$. $\blacksquare$

To eliminate link capacity violation, we revise our online routing algorithm so that a path is picked only when the capacities of all links on the path will not be violated after the new flow is admitted. This change will potentially reduce the accepted traffic ratio. As will be shown in the experiment section, the loss of accepted traffic can be well justified by completely avoiding link capacity violation.

\vspace{-5pt}\section{Evaluation}
We now evaluate the proposed models and algorithms using emerging applications over real network topologies. 

\vspace{-5pt}\subsection{Non-spilttable Flow -- VR Rendering}
\begin{figure}[htbp]
\centering
\subfigure[Starbucks Locations \label{AR:0}] {
       \includegraphics[height=1.4in,width=1.1in]{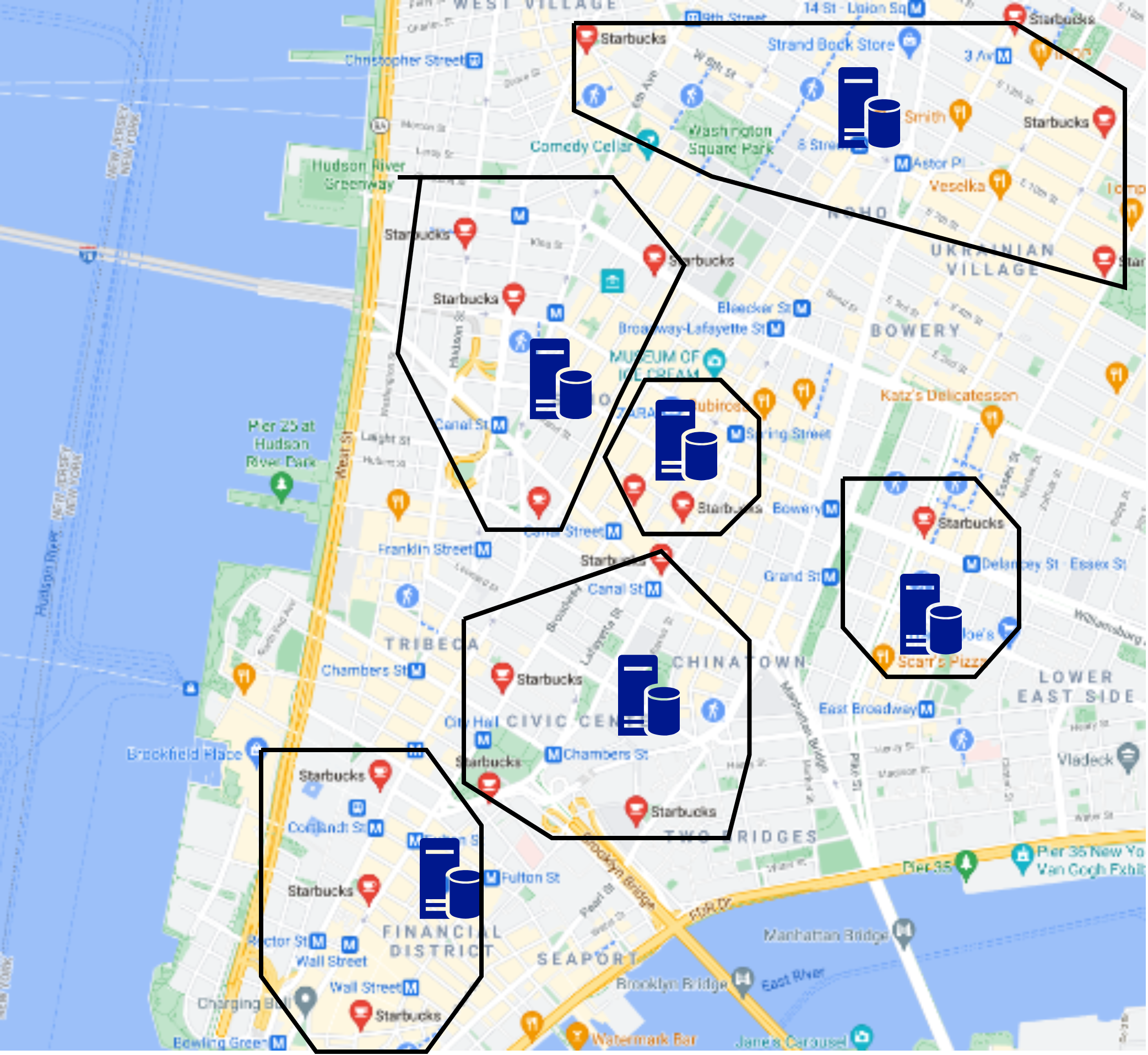}
     }
     \subfigure[Network delay for various Demand Scale \label{AR:2}] {
       \includegraphics[height=1.4in,width=1.9in]{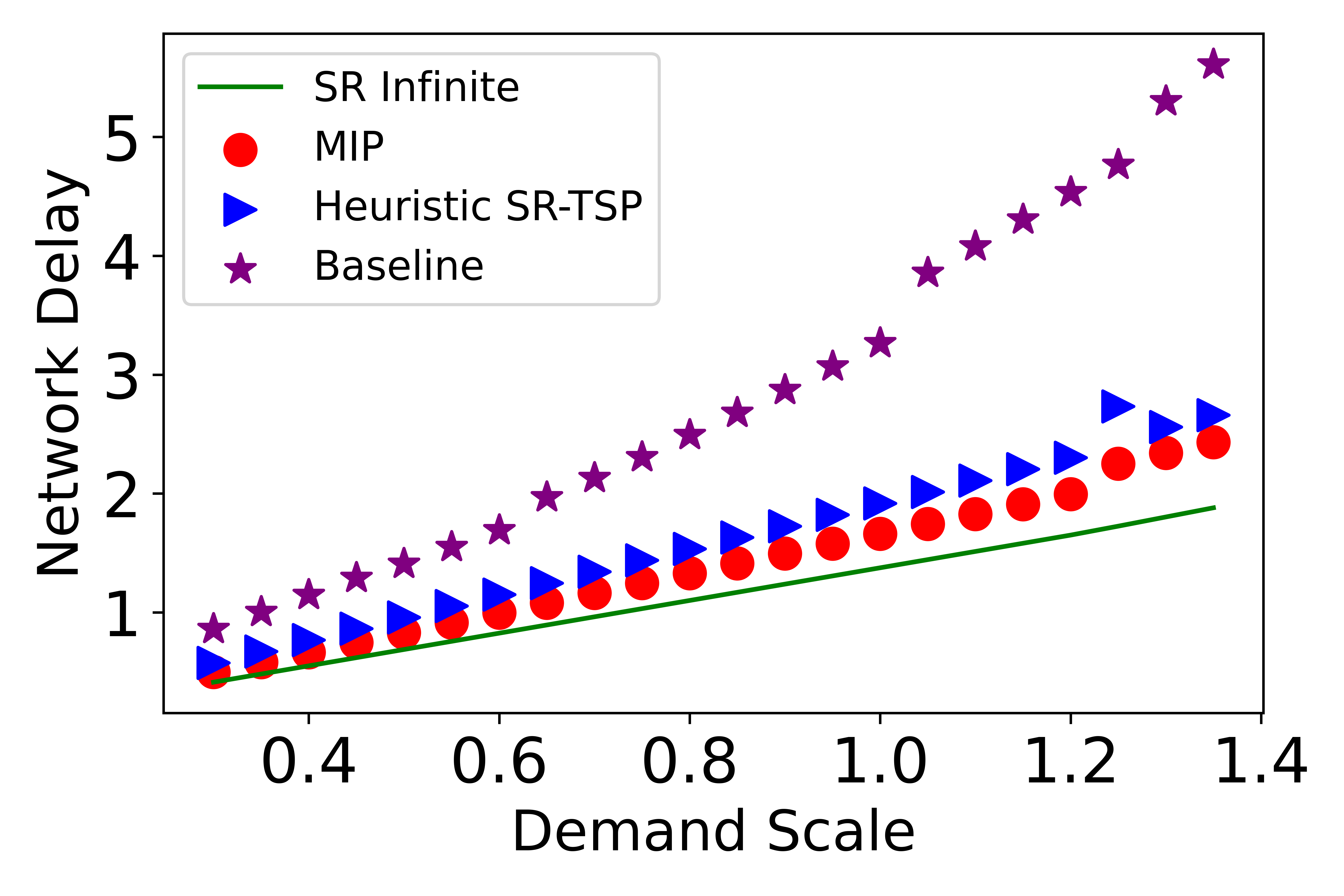}
     }
      
 \caption{Routing of Non-splittable VR Flows \label{fig:MTARoutes}}
\end{figure}

In the VR scenario, one user communicates with another user through realtime 360 degree video streaming. Rendering of 360 video is computation-intensive, it is therefore beneficial to offload the computation from the user device to edge computing nodes. The realtime video stream has to be processed as a whole, i.e., the flow is Non-Spilttable, and follows a single path. To generate a realistic VR scenario, we used the method in~\cite{wang2018service} to generate our data. For topology, we use the locations of Starbucks stores in the Lower Manhattan of New York City as the locations for users. Each store is connected to a close-by computing node, forming a star topology as shown on Fig.~\ref{AR:0}. Each computing node is connected to two or three nearby computing nodes. There are 24 nodes and 56 links in the network. We synthesize four sets of VR flows between Starbucks customers.
%without the loss of generality, the corresponding computing demand number is same as flow demand.

To evaluate the SR-TSP heuristic, in Table~\ref{AR:tb1}, we compare its solution with the optimal MIP solution on different datasets. The average performance loss of SR-TSP is about 8\% from the global optimal. In Fig.~\ref{AR:2}, we evaluate how the performance gap increases as traffic and computation demands scale up. We also develop a greedy baseline algorithm in which each source node finds the nearest computing node with sufficient computation capacity for processing, then the processed traffic is routed to the destination following the shortest path. SR-Infinite is the LP solution for infinitely splittable flow, which serves as a lower bound for the MIP solution. It can be seen that as the demands scale up, the network delay increases for all the algorithms. The gap between baseline and other algorithms is also getting larger and larger. SR-TSP uses the optimal infinitely splittable solution to get computation demand allocation and then uses mature TSP Heuristic to solve the problem of computing node traversal. It achieves a good tradeoff between time complexity and routing performance. 

\begin{table}[!ht]
\caption{Evaluation on Non-splittable VR flows~\label{AR:tb1}}
\begin{tabular}{l|l|l|l|l|l}
                & Data 1  & Data 2  & Data 3   & Data 4  & Average    \\ \hline
MIP-RINP            & 2.087  & 2.436  & 2.266   & 2.426  & 2.304  \\ \hline
SR-TSP & 2.214  & 2.589  & 2.615   & 2.622  & 2.510  \\ \hline
Gap         & 5.77\% & 5.92\% & 13.36\% & 7.48\% & 8.23\%
\end{tabular}
\end{table}

\begin{figure}[!htb]
\centering

%\subfigure[Smart City Topology\label{NYPD:1}] {
%       \includegraphics[height=1.8in,width=0.7in]{TOPO.png}
%     }
\subfigure[Network Delay for Various Spilt Scales\label{NYPD:2}] {
       \includegraphics[height=1.2in,width=1.6in]{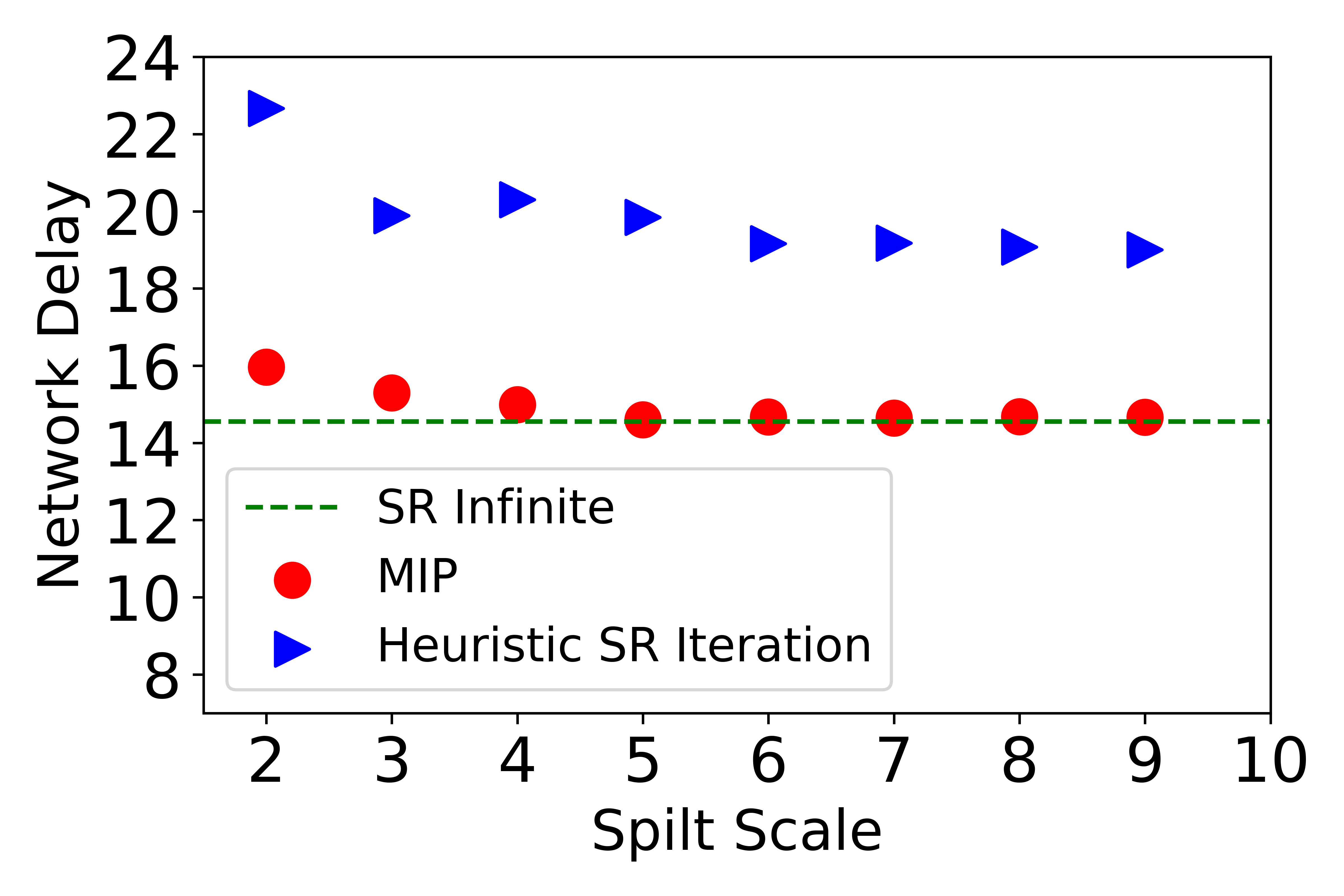}
     }
\subfigure[Number of Paths per Flow\label{NYPD:box}] {
       \includegraphics[height=1.2in,width=1.6in]{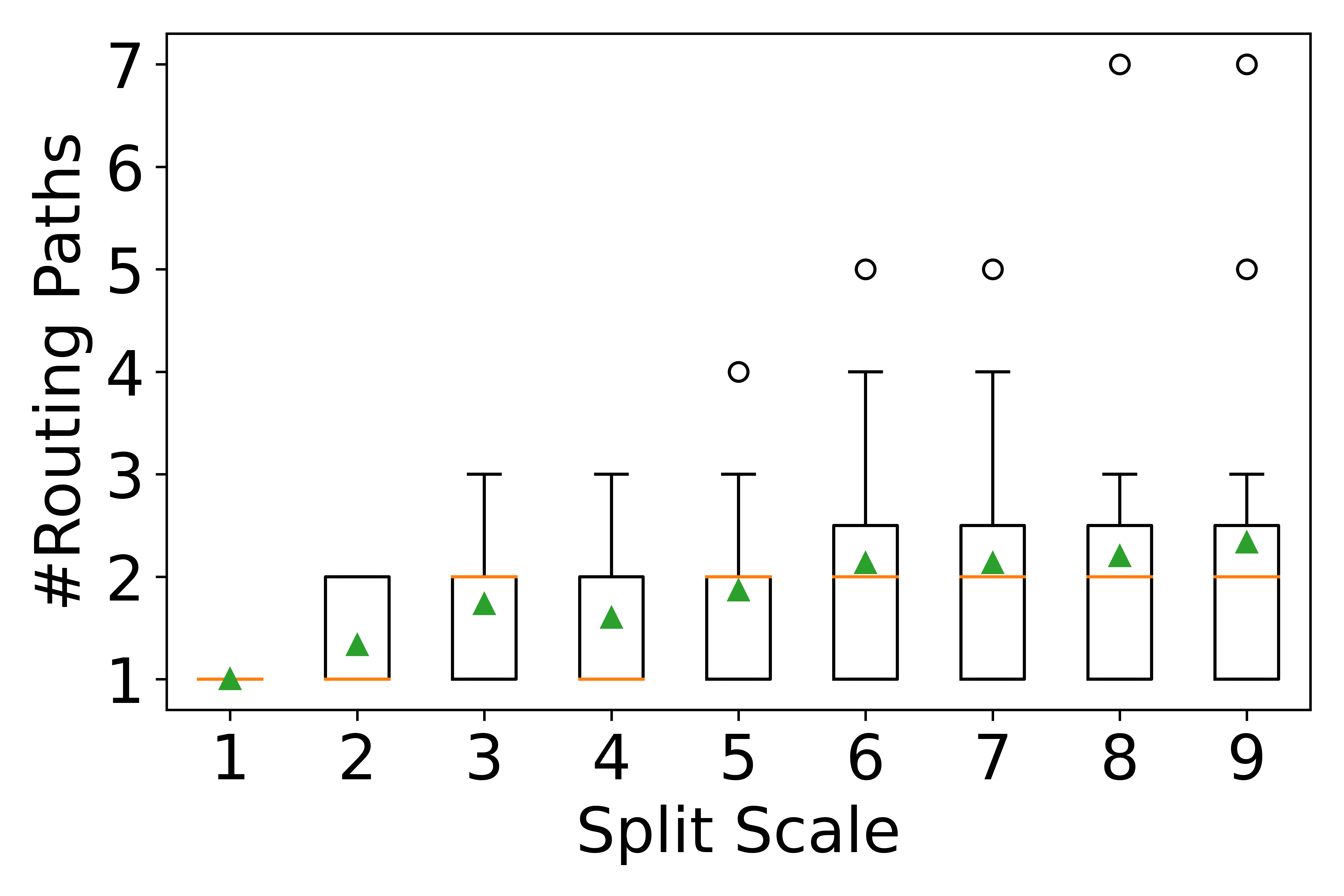}
     }
      
 \caption{K-split RINP Routing for Video Analytics Flows\label{fig:Smart}}
\end{figure}

\subsection{Finitely Splittable -- Smart City\label{smart_city}}
In the case of a smart city, there are many cameras on the streets. The rapid processing of the information captured by the cameras provides great help for the city's fire and police emergency response. %For fire alarm, proactive and rapid fire detection can reduce casualties and property losses. As for the police department, it is critical to respond quickly to crimes. For example, on the Amber system, quickly extracting vehicle plates from street camera video can save lives of kidnapped children.
Edge computing can support in-network video analytics for fast response. Unlike the non-splittable VR flows, video from one intersection is separable because there are typically multiple camera videos from different viewing angles, even video from a single viewing angle can be divided and processed in parallel. 
% 
%  This kind of real-time is not required, so there is no out of sequence problem. In video image processing, a task can be divided into multiple small parts to take different paths.

Realtime video camera data from New York City was used for evaluation. %Fig.~\ref{NYPD:1} shows the distribution of cameras from 115th to 59th Street in the Upper West Side of New York City.% 
For the same camera, there are two video analytics tasks, one for the Fire department and one for the Police. We assume that the computing power required for fire detection is proportional to the number of nearby houses. Based on the data from Zillow, we counted the number of houses within a radius of half a block. Assuming that the computing power required by the police camera is related to the traffic flow of the street, we conducted the query of the precise location traffic flow and the reasoning of the fuzzy location traffic flow according to~\cite{Traffic_Volume_Counts}\cite{okamoto2016being}\cite{NYC_Street_Camera} to complete our data set. It contains 40 nodes and 15 cameras. A total of 30 streams flow from cameras to nearby fire and police stations. 

We focus on the {\it k-split} routing. For each test, we randomly select 60\% of demands for evaluation, and the representative results are shown in Fig.~\ref{NYPD:2}. As the split scale $k$ increases, the performance gets better. For non-splitting, the performance is bad: the network delay of MIP and SR-Iteration are $245.87$ and $247.79$, respectively. But once the spilt degree goes to 2, the performance improved a lot.  SR-Infinite is the optimal solution when the splitting can go to infinity. We can see that as the $k$ goes to 4, the solution from MIP can already match the optimal. This finding is important since MIP incurs high complexity. In our example, MIP with $k=4$ uses 12.63s, but MIP with $k=9$ uses 189s, while their solution quality is almost the same. Secondly, when the network and demands are large, an alternative method is to use SR-Iteration to find a quick solution, even though it incurs $20\%$ higher delay than MIP. The solution quality of SR-Iteration also improves a lot from $k=1$ to $k=9$, but it saturates when $k$ is large. In Fig.~\ref{NYPD:box}, for the same dataset, we draw the boxplot to show how the split scale affects the number of paths taken by each flow.  The green arrow is the average number of paths taken under $k$. We can see that most flows only take less than three paths even though they are allowed to take up to $k$ paths. But some demands can still take advantage of large $k$. In the split scale $k=9$, one demand is routed to seven different paths! {\it This suggests that we can assign different splitting degrees to different demands.}

\subsection{Infinitely Splittable and Placement -- WAN}
\label{Placement}
A flow in Wide-Area-Networks (WAN) typically represents a set of user flows that share the same ingress and egress points. It is often treated as infinitely splittable in traffic engineering. We evaluate RINP with the infinitely splittable flow in the context of WAN. 

\textbf{Dataset.} We obtain two WAN datasets. The first is obtained from~\cite{uhlig2006providing}. It presents a publicly available dataset from G\'EANT, the European Research and Educational Network. 
%It provides topology information and traffic matrix per 15 minutes bin for a period of about 4 months. 
The G\'EANT topology consists of 23 nodes with 74 directed links. The average node degree is $3.217$. We randomly sample 12 node-to-node demands in the given traffic matrix. We set link capacity to 80,000 for each link and pick three nodes not in the source node set and not in the destination set as computing nodes. The second dataset is obtained by~\cite{orlowski2010sndlib}. We pick topology Abilene, which represents a high-performance backbone network across the America. It consists of 12 nodes with 30 directed links. The average node degree is 2.5. For the demand setting, we randomly sample 6 node-to-node demands in the traffic matrix. The link capacity is 40,000 for each link as described in the dataset. We also pick two nodes, SNVAng and IPLSng as computing nodes. Without loss of generality, for both datasets, we assume the computational resource demand for each flow equals its traffic volume.

\begin{figure}[!htb]
\centering
    \subfigure[G\'EANT \label{basevsseg:a}] {
       \includegraphics[height=1.06667in,width=1.6in]{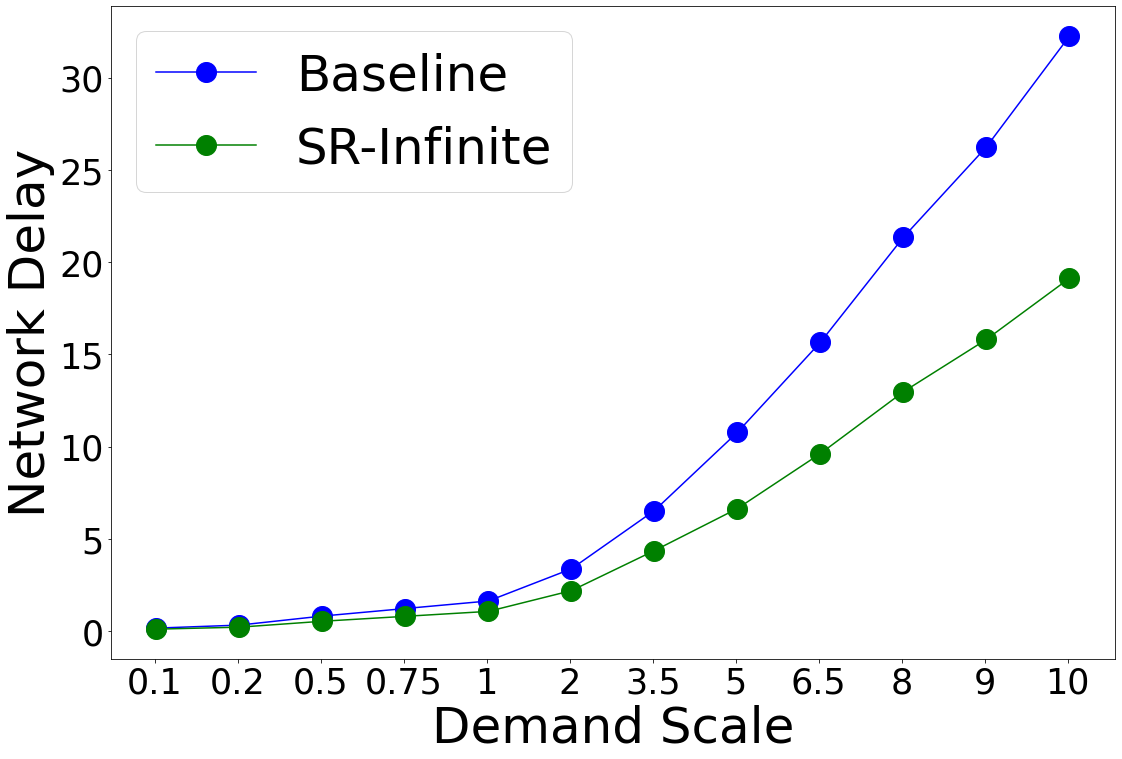}
     }
     \subfigure[Abilene \label{basevsseg:b}] {
       \includegraphics[height=1.06667in,width=1.6in]{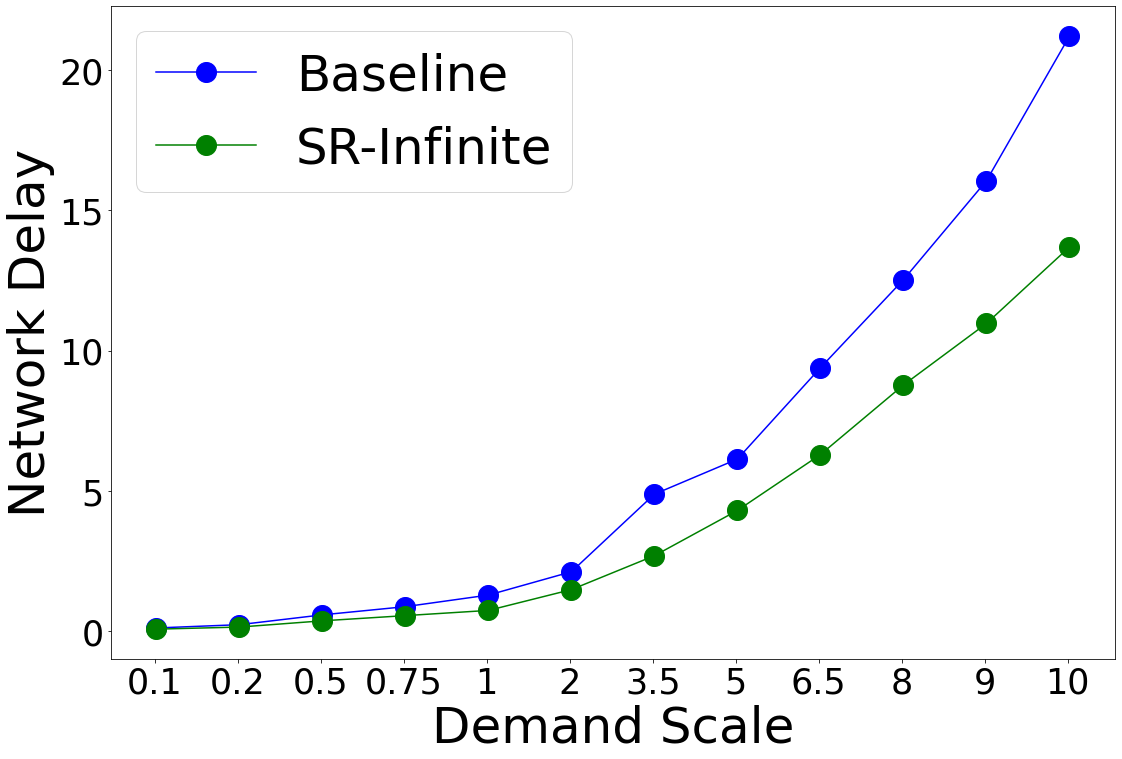}
     }
\caption{Greedy Computation Allocation v.s. Joint Optimization \label{basevsseg}}
\end{figure}

\textbf{Greedy Computation Allocation vs. Joint Optimization.} To demonstrate the importance of joint optimization of traffic routing and demand allocation, we develop a baseline algorithm that greedily allocates the flow with the largest computation load to the most powerful computing node, then conducts LP-based routing optimization for the greedy computation load allocations. The result is shown in Fig.~\ref{basevsseg}. For topology G\'EANT, the network delay of the baseline is always above the LP solution of segment routing. The average increment is about 57.58\%. For topology Abilene, the network delay of the baseline is also always above segment routing. The average increment is about 55.29\%.  

\begin{figure}
\centering
    \subfigure[G\'EANT \label{withorwithoutplacement:a}] {
       \includegraphics[height=1.06667in,width=1.6in]{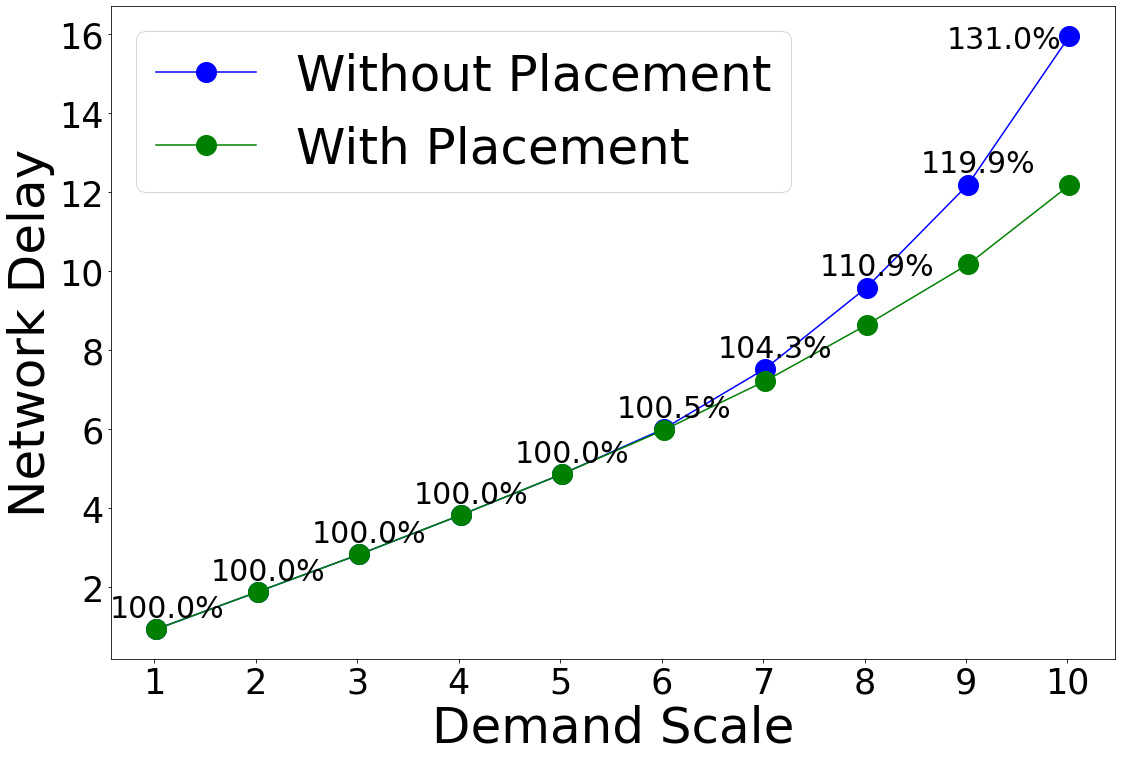}
     }
     \subfigure[Abilene \label{withorwithoutplacement:b}] {
       \includegraphics[height=1.06667in,width=1.6in]{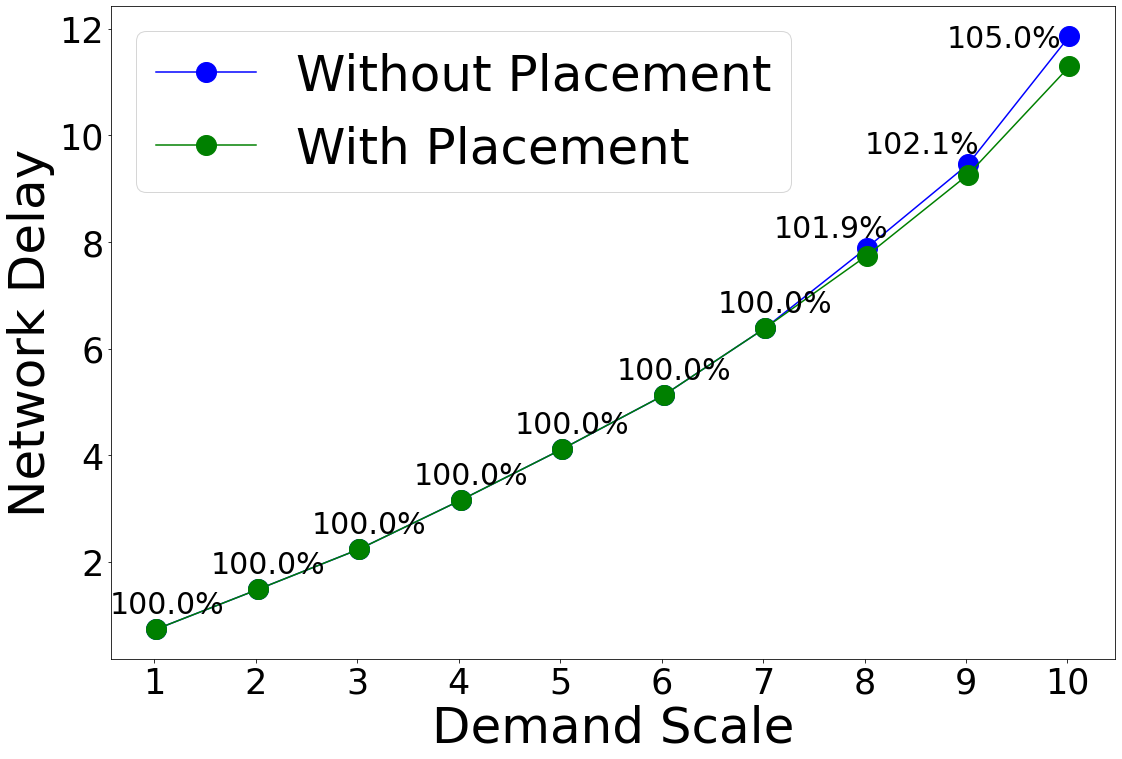}
     }
\caption{Effect of Flexible Computation Capacity Placement\label{withorwithoutplacement}}
\end{figure}

\textbf{With Placement vs. Without Placement.} We compare segment routing when the computational capacity is fixed for each computational node with when the computational capacity can be freely allocated among computational nodes as long as the sum of their computational capacity is the same as the fixed case. The result is shown in Fig.~\ref{withorwithoutplacement}. The percentage represents the ratio of the network delay without placement over that with placement. For both topologies, the network delay is the same when the demand scale is not very large. The reason is that when demand scale is small, the original fixed computational capacity distribution is enough to handle all demands that follow the best routes. For topology G\'EANT, with placement version outperforms the fixed version when the demand scale goes up to 6. When demand scale is 10, the network delay of the fixed version is about 131.0\% of the with placement version. For topology Abilene, with placement version outperforms the fixed version when the demand scale goes up to 8. When the demand scale is 10, the network delay of the fixed version is about 105.0\% of the network delay of the with placement version.

\begin{figure}
\centering
    \subfigure[Initial Topology \label{placementdifftopo1:a}] {
       \includegraphics[height=1in,width=1in]{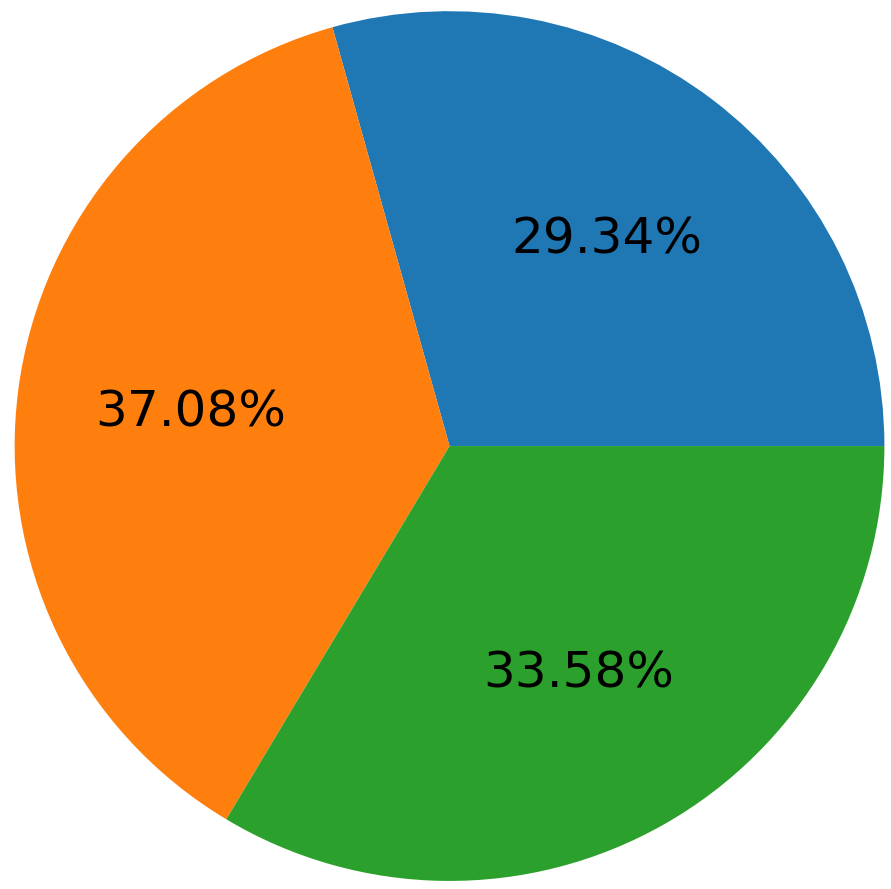}
     }
     \subfigure[Cut Three Links \label{placementdifftopo1:b}] {
       \includegraphics[height=1in,width=1in]{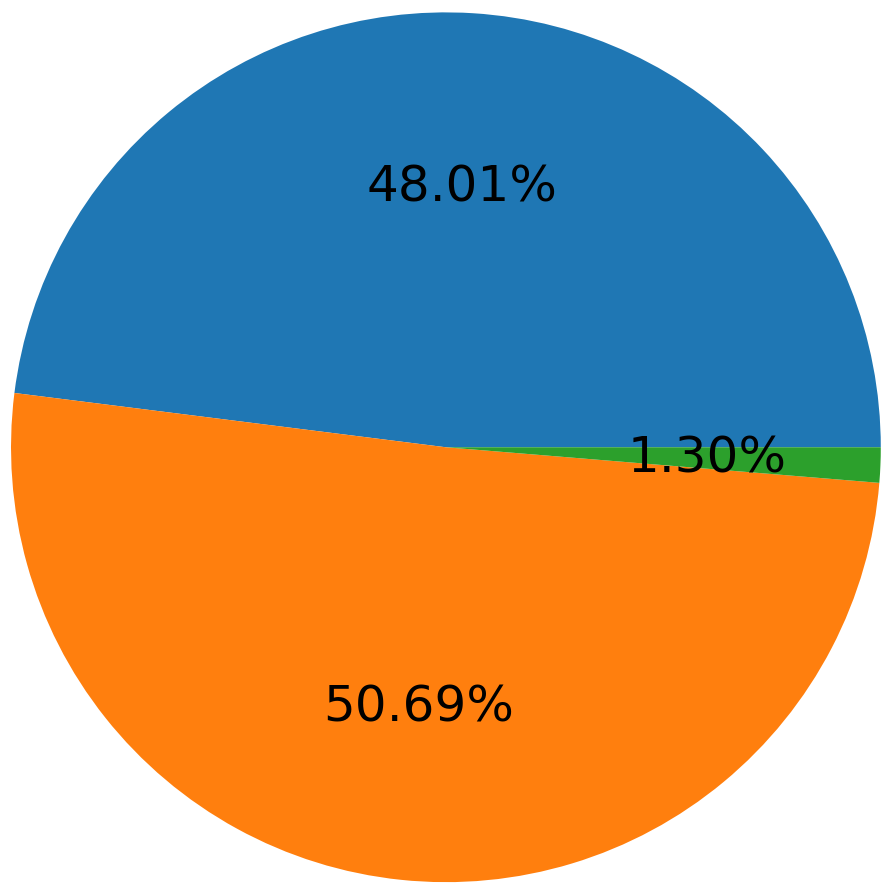}
     }
     \subfigure[Cut Five Links \label{placementdifftopo1:c}] {
       \includegraphics[height=1in,width=1in]{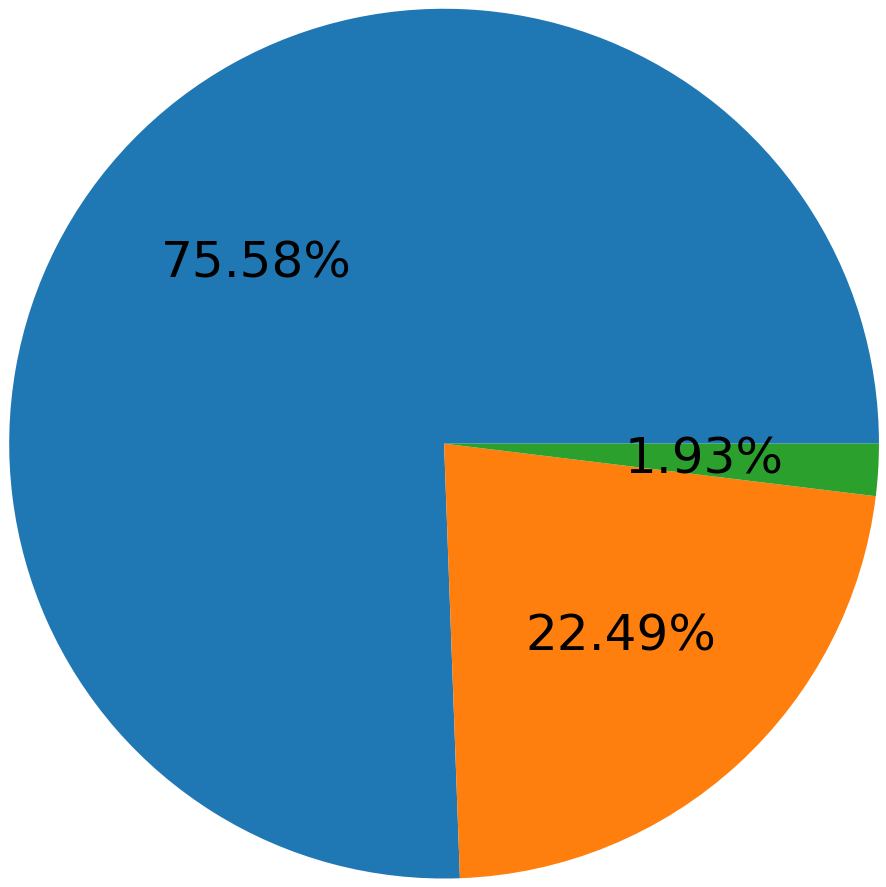}
     }
\caption{Capacity Placement Change after Link Failures for G\'EANT \label{placementdifftopo1}}
\end{figure}

\textbf{Adaptation to Link Failures.} We run segment routing when computational capacity can be freely allocated among computational nodes under a given budget. For both topologies, we run it on three different configurations by disabling some links. The pie chart shows the ratio of the computational capacity allocated to each computation node. The result for G\'EANT is shown in Fig.~\ref{placementdifftopo1}. Blue represents node 3, orange represents node 13, and the green represents node 16. Initially, the allocation of the computational capacity among node 3, 13 and 16 is shown as Fig.~\ref{placementdifftopo1:a}. After cutting three bidirectional links connected to node 16, the allocation is shown as Fig.~\ref{placementdifftopo1:b}. The computational capacity allocated to node 16 is drastically reduced. After cutting two more bidirectional links connected to node 13, the allocation is shown in Fig.~\ref{placementdifftopo1:c}. The computational capacity allocated to node 13 also drastically reduces, and most computational capacity is allocated to node 3. Similar results for Abilene network can be found in our technical report~\cite{Lifan_Report22}.

\begin{figure}
\centering
    \subfigure[G\'EANT \label{considerscale:a}] {
       \includegraphics[height=1.06667in,width=1.6in]{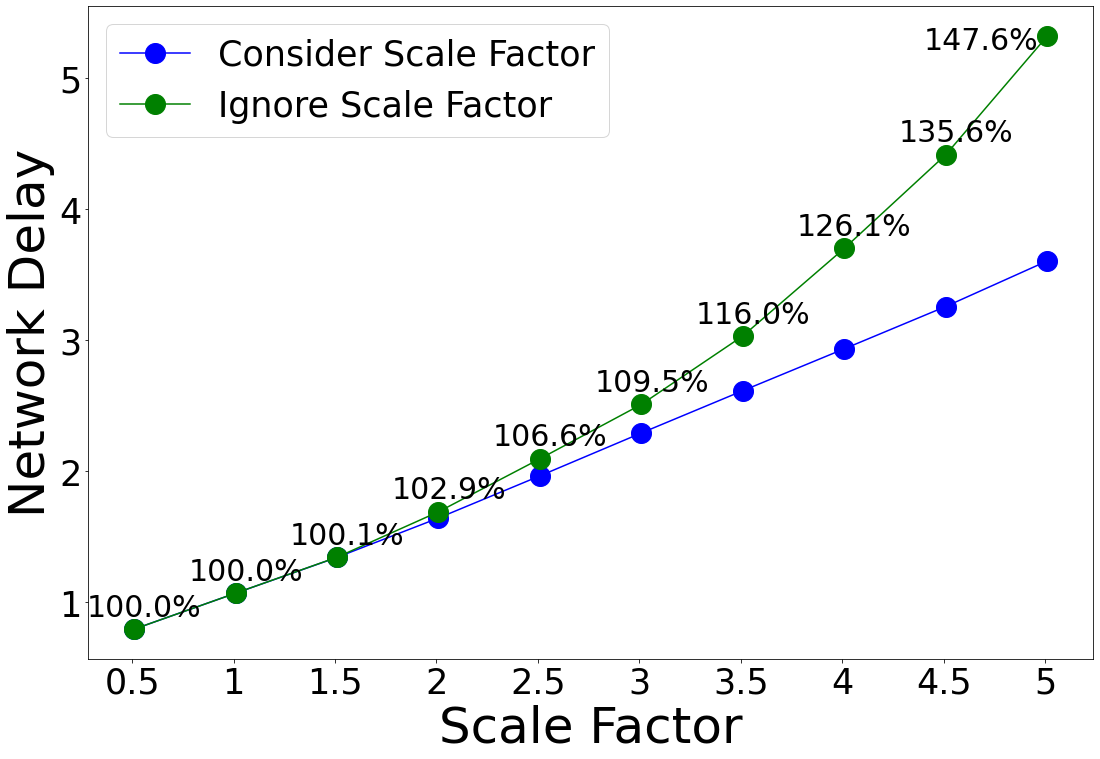}
     }
     \subfigure[Abilene \label{considerscale:b}] {
       \includegraphics[height=1.06667in,width=1.6in]{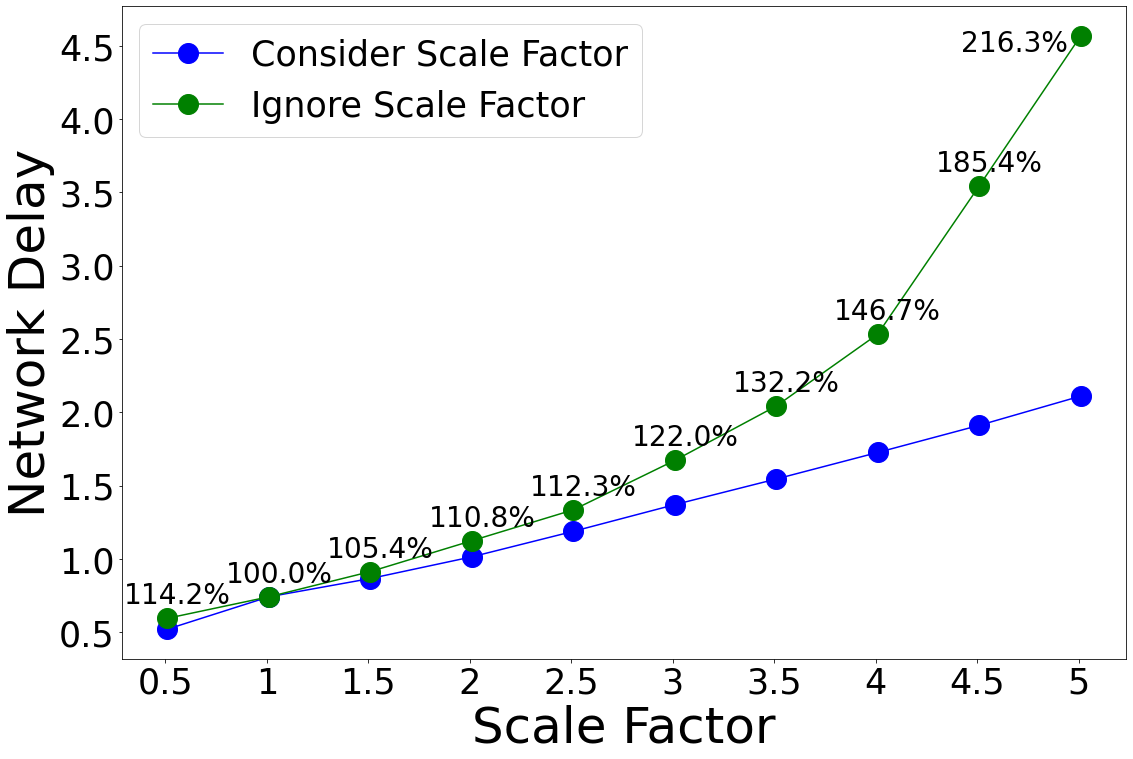}
     }
\caption{Consider Traffic Scaling vs. Ignore Traffic Scaling \label{considerscale}}
\end{figure}

\textbf{Impact of Traffic Scaling after Processing.} In this part, we run segment routing when the flow size changes after going through a computational node. We first take the scale factor into account in our model for different scale factors. We then compare it with the case when the traffic scaling is ignored when calculating the optimal segment routing. The result is shown in Fig.~\ref{considerscale}. The percentage represents the ratio of the network delay when ignoring traffic scaling to the network delay when considering traffic scaling. For topology G\'EANT, the network delay is shown in Fig.~\ref{considerscale:a}. Except for the scale factor 0.5 and 1, the network delay of ignoring the scale factor is higher than that of considering the scale factor. For topology Abilene, the network delay is shown in Fig.~\ref{considerscale:b}. Except the scale factor 1, the network delay of ignoring the scale factor is higher than that of considering the scale factor.

\subsection{Online Routing Algorithm with Guarantee}

\textbf{Dataset.}
To evaluate the online routing algorithm, we synthesize dynamic traffic demands. The topology is the same as the Smart City problem in Section~\ref{smart_city}. In the graph of 40 nodes, 14 nodes were selected as computing nodes. Without loss of generality, the ratio between the computing demand and the flow size is $1:1$. Each link capacity is 550 units, and node capacity is 300 units. We pick 8 source-destination pairs. New flows arrive at the network according to the Poisson process and are assigned to each source-destination pair in a round-robin fashion. The time duration of each flow follows lognormal distribution, i.e., $\tau=e^{\mu+\sigma Z}$, where $Z$ is the standard normal variable. When a flow is active, the traffic volume follows the Gaussian distribution.

\begin{figure}[!htb]
\centering
       \includegraphics[height=1.4in,width=2.1in]{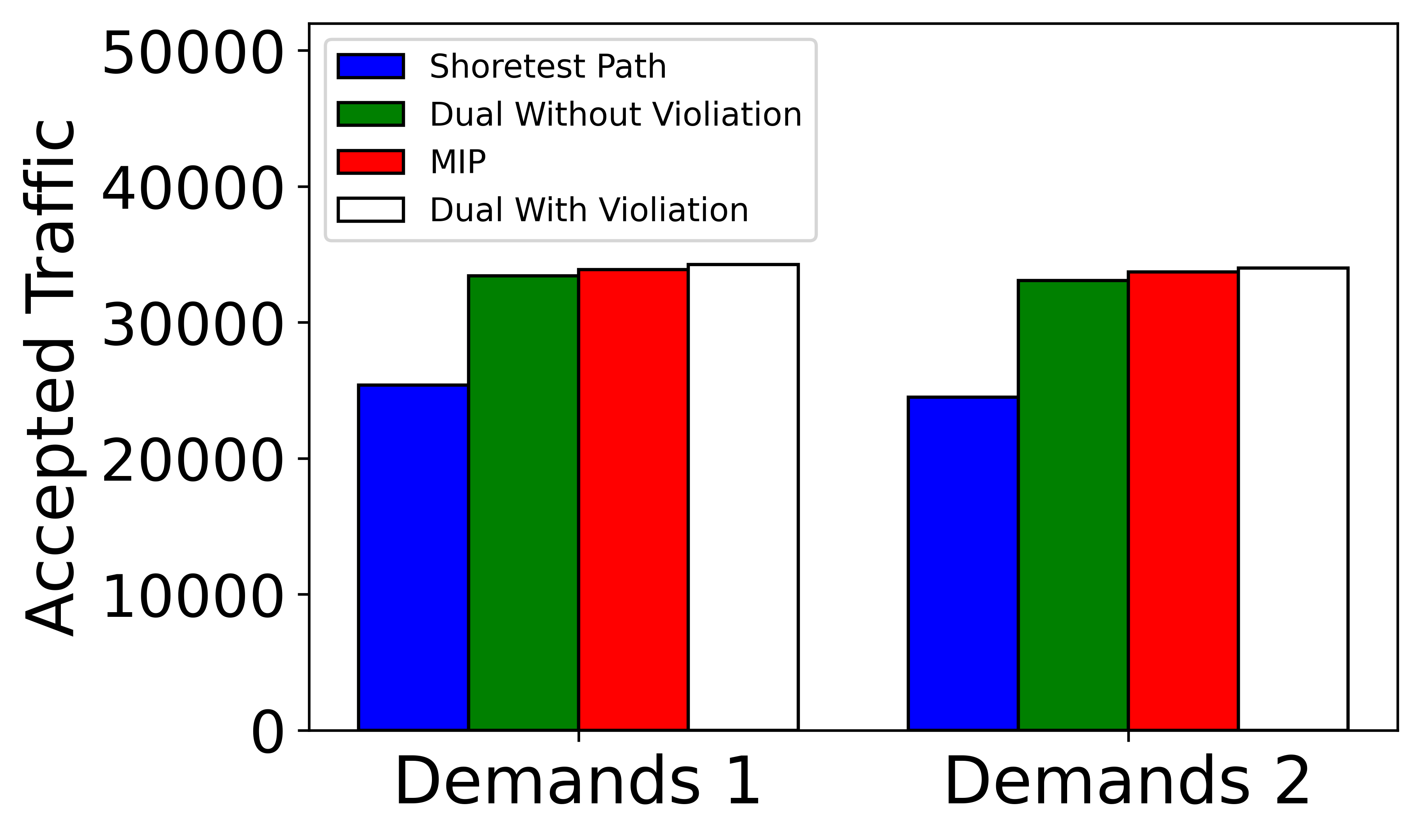}
\caption{Total Accepted Traffic of Dynamic Flows\label{P3:perf}}
\end{figure}

\begin{figure}[!htb]
   \begin{minipage}[b]{0.48\linewidth}
     \centering
     \includegraphics[height=1in,width=1.5in]{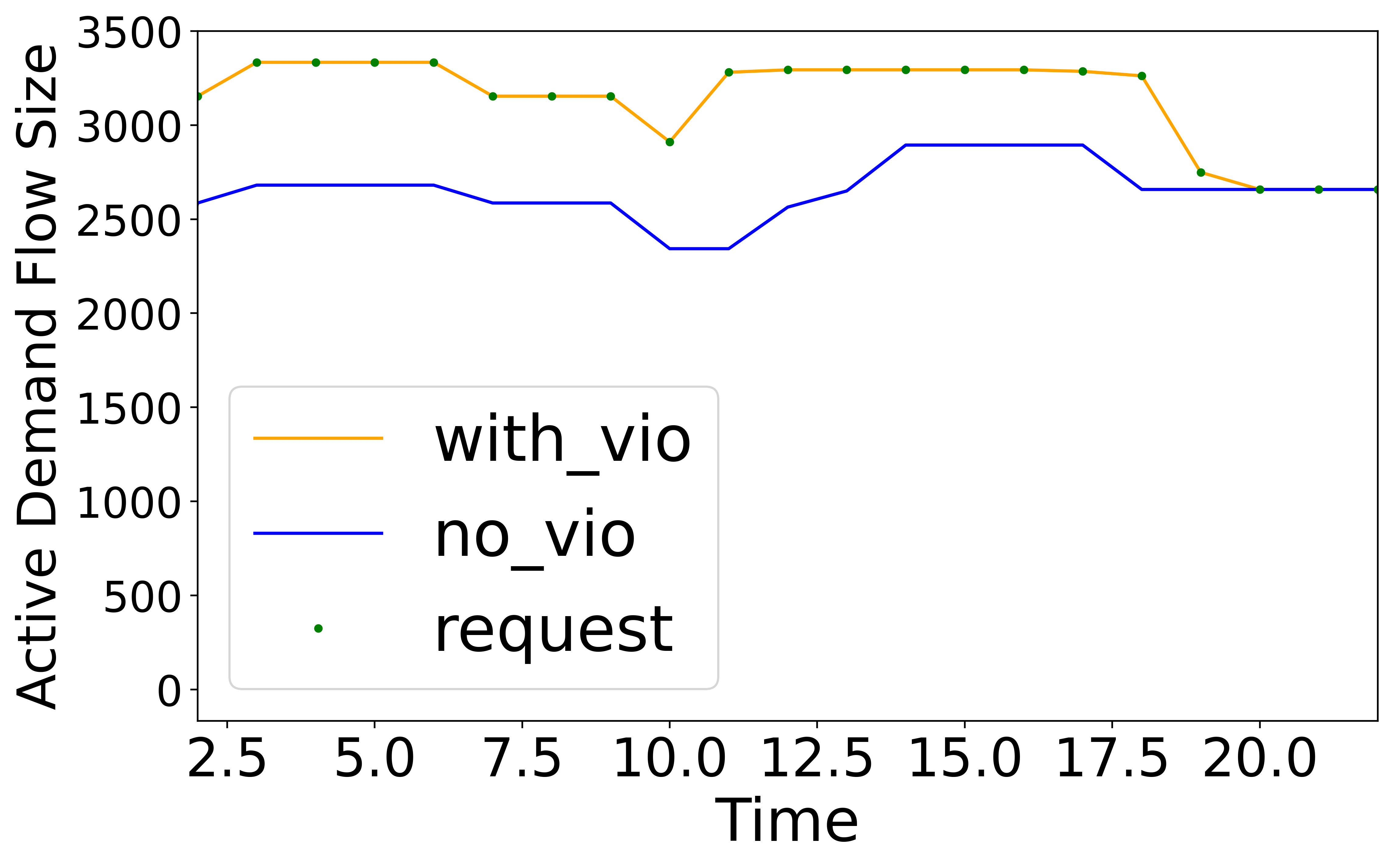}
     \caption{Active Flow Size Over Time\label{P3:active flow over time}}
   \end{minipage}
   %\hfill
   \begin{minipage}[b]{0.48\linewidth}
     \centering
     \includegraphics[height=1in,width=1.5in]{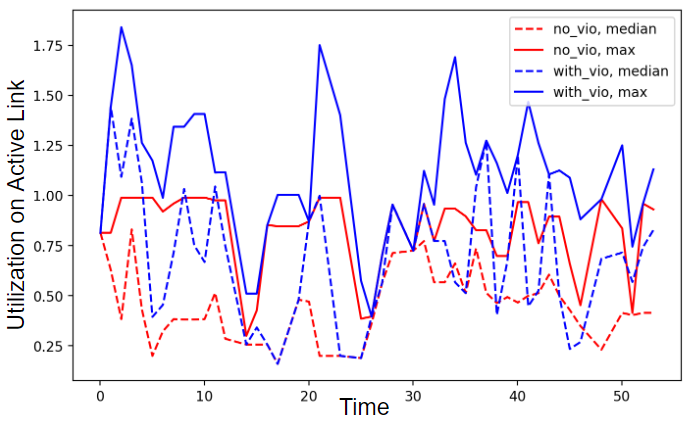}
     \caption{Link Violation Over Time\label{Median Max Link By Time}}
   \end{minipage}
\end{figure}

\textbf{Performance.}
For the online algorithm, we compare two versions, one with link violation and one without violation. We also use the shortest path algorithm as the baseline, and the upper bound is obtained from the MIP model implemented by Gurobi. In  Fig.~\ref{P3:perf}, the flow arrival rate is two flows/minute, the lognormal duration parameters are $\mu=0.974$ minutes, and $\sigma=0.5$ minutes. The average flow volume is $85$ units. To evaluate the impact of flow volume variations, we conducted two experiments with flow volume standard deviation of $10$ and $20$, respectively. Our online algorithms significantly outperform the shortest path routing. The accepted traffic of the online algorithm without violation is only slightly lower than the MIP solution. With link violation allowed, the online algorithm can accept a little bit more traffic than MIP. Fig.~\ref{P3:active flow over time} shows the total traffic of active flows over time. This suggests that our online algorithm can efficiently utilize available communication and computation resources in the network to support dynamic flows.  

\begin{figure}[!htb]
\centering
       \includegraphics[height=1.5in]{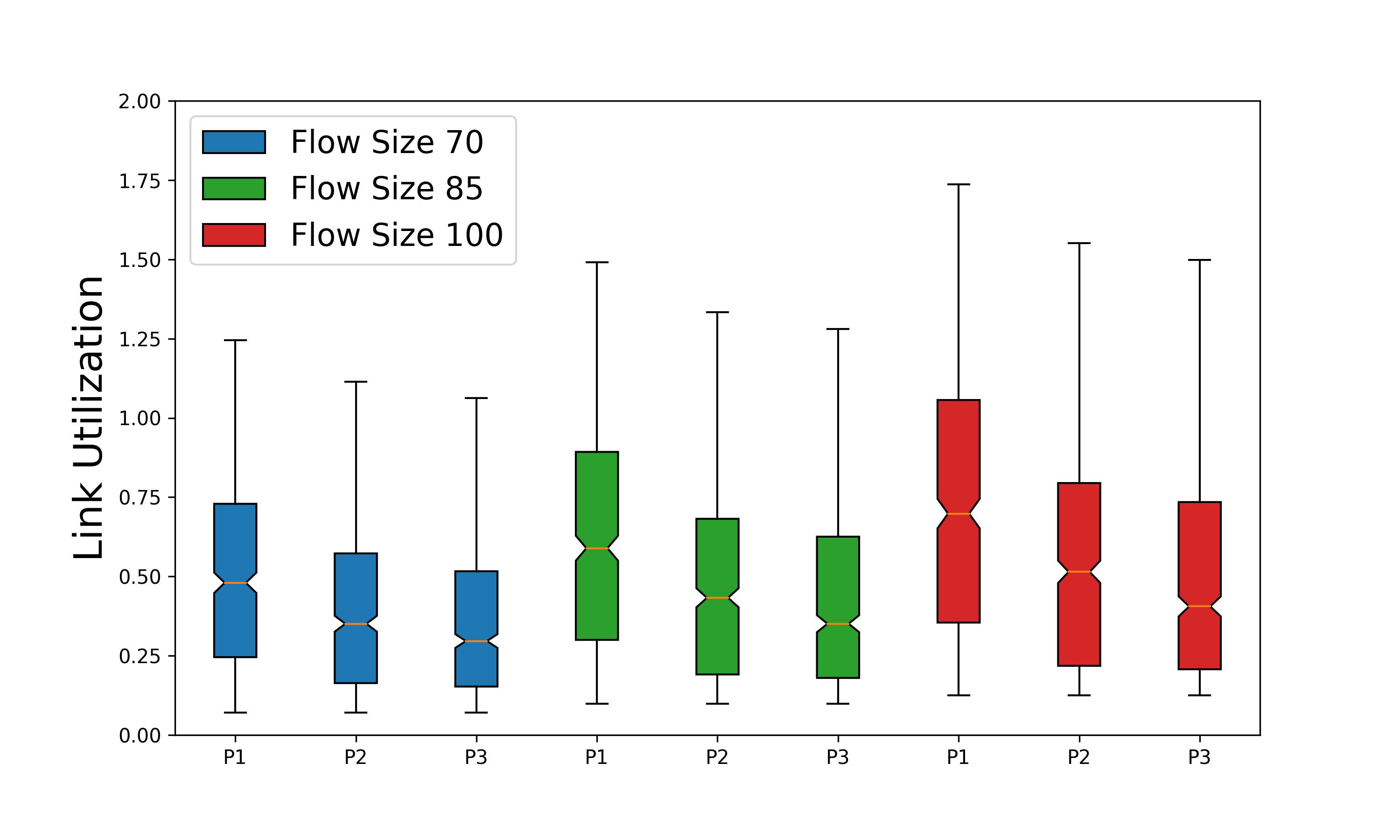}
\caption{Link Utilization for Online Routing with  Violation\label{vio_diff_para}}
\end{figure}

\begin{figure}[!htb]
\centering
       \includegraphics[height=1.5in]{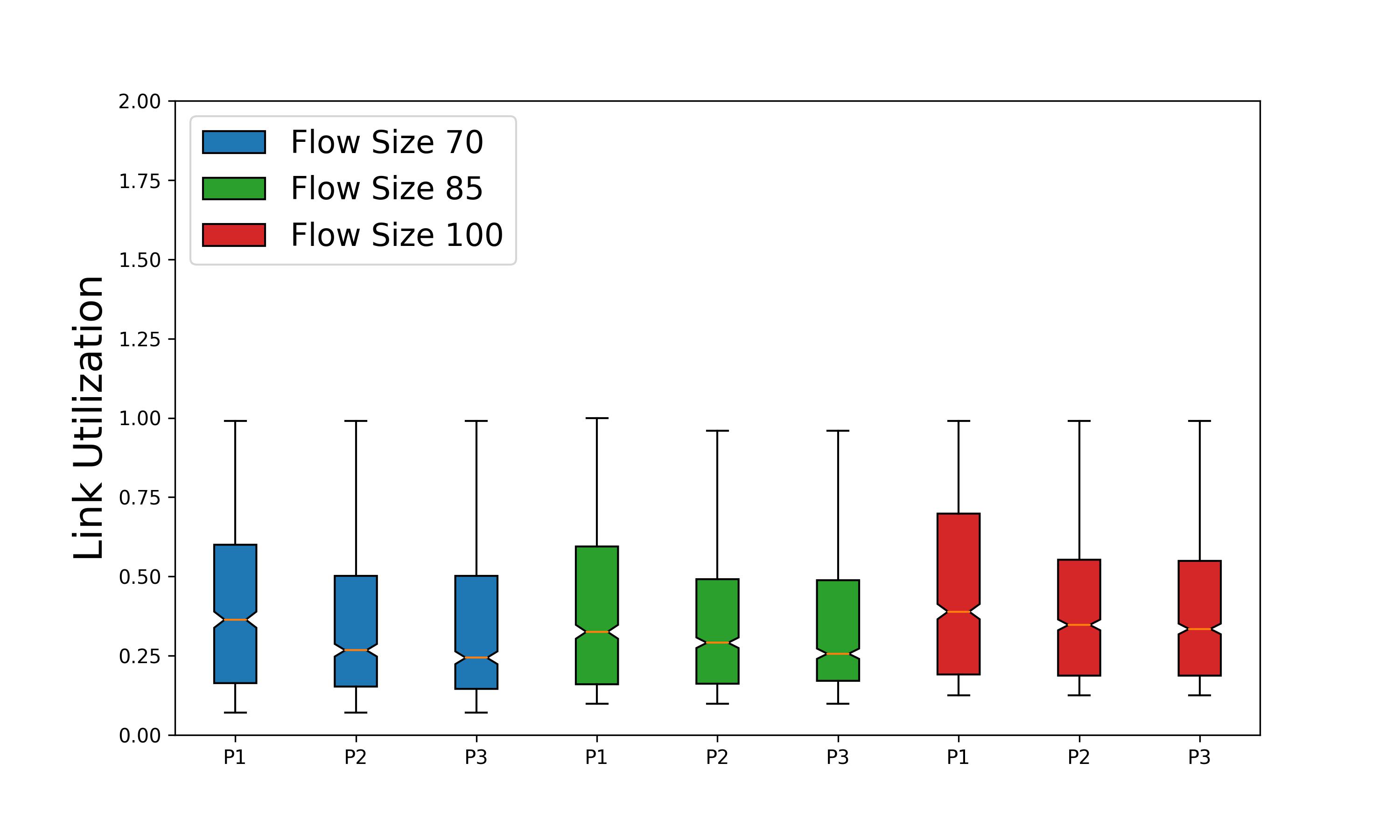}
\caption{Link Utilization for Online Routing without  Violation\label{novio_diff_para}}
\end{figure}

\textbf{Violation.}
In the online routing algorithm, the link violation theoretical upper bound has been given. But in practice, within a reasonable demand flow size range, the violation is rare. To check the actual violation, Fig.~\ref{Median Max Link By Time} plots the Median and Max link utilization over time. We also plot the boxplot distribution of link utilization in Fig.~\ref{vio_diff_para} and Fig.~\ref{novio_diff_para} for different demand parameter settings. For all the experiments, the flow arrival rate is still two flows/minute. There are three settings of flow time duration parameters $(\mu,\sigma)$: $(0.974, 0.5)$, $(0.598, 1.0)$, $(0.379, 1.2)$, respectively. Under those settings for lognormal distribution, the average duration is the same, but the variance increases from the first setting to the last setting. We also try three average flow sizes of $70$, $85$, and $100$, with the standard deviation fixed at $10$. It is worth noting that with flow size expectation of $85$, the network capacity is reached~\footnote{For each demand pair, we calculate its long-term average traffic volume based on the average flow arrival rate, mean flow duration, and mean flow size. We then check the feasibility of the long-term average traffic matrix  of all pairs by solving the static LP routing optimization.}. In Fig.~\ref{vio_diff_para}, as the average flow size increases, the link utilization increases. With the same average flow size, as the duration variance increases, the link utilization decreases, this is because more dynamic flow duration leads to more rejected flows by the online routing algorithm. Even though the maximum utilization can go up to $1.75$, utilization on most of the links are still below one. In Fig.~\ref{novio_diff_para}, with capacity  violation control, no utilization goes over $1$, and the differences between different flow sizes and different duration variations become smaller, while the overall trends agree with Fig.~\ref{vio_diff_para}.

\vspace{-0.25in}

\vspace{-0.2cm}

\section{Conclusion}
\label{sec:conclusion}
%\vspace{-0.2cm}
In this paper, we studied optimal routing in networks with embedded computational services. We developed routing optimization models and fast heuristic algorithms that take into account the unique features of routing with in-network processing and various routing requirements resulting from the computation needs of diverse applications. For the dynamic demand scenario, we came up with an online routing algorithm with a performance guarantee. We demonstrated through evaluations that our models and algorithms are highly customizable and can achieve close-to-optimal performance in a wide range of application scenarios. While the current work is focused on optimizing the network performance, we will investigate the robustness and resilience of RINP in our future work. In particular, we are interested in exploring how traffic routing and computation resource provisioning can help each other to quickly recover from major failures in networks with embedded computational services, as demonstrated in our preliminary results in Section~\ref{Placement}.

\section*{Acknowledgement}
This work was partially supported by USA National Science Foundation under contract CNS-2148309. 

\bibliographystyle{IEEEtran}
\bibliography{TON_20220815.bib}
%\bibliography{INFO_bst.bib}

%\begin{thebibliography}{00}

%\bibitem{mei2019realtime}
%Mei, L., Hu, R., Cao, H., Liu, Y., Han, Z., Li, F. and Li, J., 2019, March. %Realtime mobile bandwidth prediction using lstm neural network. In International Conference on Passive and Active Network Measurement (pp. 34-47). Springer, Cham.

%\bibitem{xu2019joint}
%Xu, Xiaolong, Chengxun He, Zhanyang Xu, Lianyong Qi, Shaohua Wan, and Md Zakirul Alam Bhuiyan. "Joint optimization of offloading utility and privacy for edge computing enabled IoT." IEEE Internet of Things Journal 7, no. 4 (2019): 2622-2629.

%\bibitem{GEANTopology}
%S. Uhlig, B. Quoitin, S. Balon and J. Lepropre. Providing public intradomain traffic matrices to the research community. ACM SIGCOMM Computer Communication Review, 36(1), January 2006.

%ordered

%\end{thebibliography}

\end{document}